\documentclass[twocolumn,showpacs,preprintnumbers,amsmath,amssymb, prb]{revtex4}
\usepackage{graphicx}% Include figure files
\usepackage{dcolumn}% Align table columns on decimal point
\usepackage{bm}% bold math
\begin{document}

\preprint{APS/123-QED}
\title{Kondo lattice and antiferromagnetic behavior in quaternary\\ CeTAl$_4$Si$_2$ (T~=~Rh, Ir) single crystals}

\author{Arvind Maurya, R. Kulkarni, A. Thamizhavel}
\affiliation{Department of Condensed Matter Physics and Materials
Science, Tata Institute of Fundamental Research, Homi Bhabha Road,
Colaba, Mumbai 400 005, India.}

\author{D. Paudyal}
\affiliation{The Ames Laboratory, U.S. Department of Energy, Iowa State University, Ames, Iowa 50011-3020, USA.}

\author{S. K. Dhar}
\email{sudesh@tifr.res.in}
\affiliation{Department of Condensed Matter Physics and Materials
Science, Tata Institute of Fundamental Research, Homi Bhabha Road,
Colaba, Mumbai 400 005, India.}

\date{\today}% It is always \today, today,
%  but any date may be explicitly specified

\begin{abstract}
We report the synthesis and the magnetic properties of single crystalline CeRhAl$_4$Si$_2$ and CeIrAl$_4$Si$_2$ and their non magnetic La-analogs.  The single crystals of these quaternary compounds were grown using Al-Si binary eutectic as flux. The  anisotropic magnetic properties of the cerium compounds were explored in detail by means of magnetic susceptibility, isothermal magnetization, electrical resistivity at ambient and applied pressures up to 12.6~kbar, magnetoresistivity and heat capacity measurements. Both CeRhAl$_4$Si$_2$ and CeIrAl$_4$Si$_2$ undergo two antiferromagnetic transitions, first from the paramagnetic to an antiferromagnetic state at $T_{\rm N1}$~=~12.6~K and 15.5~K, followed by a second transition at lower temperatures $T_{\rm N2}$~=~9.4~K and 13.8~K (inferred from the peaks in the heat capacity), respectively, in conformity with an earlier report in the literature. The paramagnetic susceptibility is highly anisotropic and its temperature dependence in the magnetically ordered state suggests the \textbf{c}\--axis to be the relatively easy axis of magnetization. Concomitantly, isothermal magnetization at 2~K along the \textbf{c}\--axis shows a sharp spin-flop transition accompanied by a sizeable hysteresis, while it varies nearly linearly with field along the [100] direction up to the highest field 14~T of our measurement. The electrical resistivity provides evidence of the Kondo interaction in both compounds, inferred from its $-lnT$ behavior in the paramagnetic region and the decrease of magnetic transition temperature with pressure. The heat capacity data confirm the bulk nature of the two magnetic transitions in each compound, and further support the presence of Kondo interaction by a reduced value of the entropy associated with the magnetic ordering. From the heat capacity data below 1~K, the coefficient of the linear term in the electronic heat capacity, $\gamma$, is inferred to be 195.6 and 49.4~mJ/mol K$^2$ in CeRhAl$_4$Si$_2$ and CeIrAl$_4$Si$_2$, respectively classifying these materials as moderate heavy fermion compounds. The main features of the magnetoresistivity measured at a particular temperature correlate nicely with the isothermal magnetization at the same temperature in these two isostructural compounds.  We have also carried out an analysis of the magnetization based on the point charge crystal electric field model and derived the crystal electric field energy levels which reproduce fairly well the peak seen in the Schottky heat capacity in the paramagnetic region. Further, we have also performed electronic structure calculations using (LSDA +U) approach, which provide physical insights on the observed magnetic behaviour of these two compounds.

\end{abstract}

% PACS, the Physics and Astronomy
% Classification Scheme.

%Use showkeys class option if keyword display desired

\maketitle
\section{Introduction}
Recently, the synthesis and magnetic properties of quaternary EuTAl$_4$Si$_2$ (T~=~Rh and Ir) single crystals, using the Al-Si binary eutectic as flux have been reported~\cite{Maurya, Maurya2}. The two Eu compounds initially order into an incommensurate amplitude modulated antiferromagnetic state at $T_{\rm N1}$~=~11.7 and 14.7~K respectively, followed by a second transition to an equal moment state at lower temperature $T_{\rm N2}$. Though these two compounds prima-facie are antiferromagnetic, the isothermal magnetization curves at low temperatures (below $T_{\rm N2}$) show a hysteresis right near the origin with a remnance; unlike any other antiferromagnetic material. The EuTAl$_4$Si$_2$ compounds adopt an ordered derivative of the ternary KCu$_4$S$_3$-type tetragonal, $tP8$, $P4/mmm$ structure, which leads to quaternary and truly stoichiometric 1:1:4:2 compounds, with a local fourfold axial ($4/mmm$) symmetry at the Eu site.  These are a new addition to several quaternary rare earth-based compounds already known in the literature with 1:1:4:2 stoichiometry, which have been grown using aluminum as flux; for example, RNiAl$_4$(Ni$_x$Si$_{2-x}$), EuNiAl$_4$Si$_2$, RNiAl$_4$Ge$_2$, RAuAl$_4$Ge$_2$ and RAuAl$_4$(Au$_x$Ge$_{1-x}$)$_2$ where R is a rare earth metal~\cite{Wu,Latturner}. While most of the these compounds adopt the rhombohedral YNiAl$_4$Ge$_2$-type structure~\cite{Sieve}, the phases RAuAl$_4$(Au$_x$Ge$_{1-x}$)$_2$ and EuAu$_{1.95}$Al$_4$Ge$_{1.05}$ crystallize in the KCu$_4$S$_3$-type structure~\cite{Sieve}. Both structure types are characterized by the slabs of ``AuAl$_4$X$_2$ (X~=~Si or Ge)" or ``AuAl$_4$(Au$_x$Ge$_{1-x}$)$_2$" stacked along the \textbf{c}\--axis with layers of R atoms in between. The Ce atoms in both CeAuAl$_4$Ge$_2$ and CeAuAl$_4$(Au$_x$Ge$_{1-x}$)$_2$ (x~=~0.4) have been reported to be in a valence fluctuating state~\cite{Wu}. This suggests the possibility of strong hybridization between the Ce-$4f$-orbitals and the itinerant electrons in these structure-types, which is known to lead to a variety of anomalous ground states, such as Kondo lattices, heavy fermions with huge effective electron masses, magnetically ordered states with reduced saturation moments~\cite{Stewart, Steglich}. The N\'eel temperature in some heavy fermion, antiferromagnetic Kondo lattices can be tuned to zero using pressure as an external parameter, which leads to a quantum phase transition where the Fermi-Landau description of quasiparticles breaks down~\cite{Stewart2}. It was therefore of interest to explore the formation of other RTAl$_4$Si$_2$ (T~=~Rh and Ir) compounds, in particular for R~=~Ce. We have been able to grow the single crystals for R~=~Ce and Pr and in this report we give a detailed description of properties of two Ce compounds, using the techniques of magnetization, electrical resistivity  in zero and applied magnetic fields, and under externally applied pressure, and heat capacity. We find that both CeTAl$_4$Si$_2$ (T~=~Rh and Ir) compounds are dense Kondo lattice antiferromagnets, each undergoing two magnetic transitions like the Eu-analogs. We have supplemented our experimental data with electronic structure calculations for CeTAl$_4$Si$_2$ (T = Rh, Ir and Pt) employing the local spin density approximation including Hubbard $U$\--onsite electron-electron correlation. 

While this manuscript was under preparation, Ghimire \textit{et al.}~\cite{Ghimire} have reported the anisotropic susceptibility and resistivity, and the low temperature heat capacity (1.8 to 30 K) of CeMAl$_4$Si$_2$ (M = Rh, Ir, Pt) compounds~\cite{Ghimire}. Our observation of two antiferromagnetic transitions in these two compounds (M~=~Rh and Ir) is in conformity with their results. Ghimire~\textit{et al.} allude to the presence of Kondo interaction and the opportunity to explore strongly correlated electron behaviour in these quaternary compounds. Our low temperature heat capacity data (below 1~K) reveal moderately heavy electron masses.  Our more extensive magnetization data, and magnetoresistivity reveal some additional features in the properties of the these two compounds. Our resistivity data at ambient pressure show the presence of Kondo interaction in these compounds, further supported by the resistivity data under externally applied pressure, where we observe a suppression of $T_{\rm N}$ with pressure, which is a standard feature of Ce-Kondo lattices predicted by the Doniach's phase diagram~\cite{Doniach}.
In a more recent paper, using powder neutron diffraction, Ghimire \textit{et al.}~\cite{Ghimire_Neutron} have determined that the antiferromagnetic structure in both CeRhAl$_4$Si$_2$ and CeIrAl$_4$Si$_2$ is A-type with propagation vector $\textbf{k}=(0,0,1/2)$ along the \textbf{c}\--axis.

\section{Experiment}
 The single crystals of CeTAl$_4$Si$_2$ and LaTAl$_4$Si$_2$ were grown by following the same experimental protocol as described in ref.~\onlinecite{Maurya} for the Eu-compounds. Their Laue patterns were recorded using a Huber Laue diffractometer, while the phase purity was inferred from the powder x-ray diffraction pattern collected using a PANalytical x-ray diffractometer. The stoichiometry was checked by semiquantitative analysis performed by energy dispersive analysis by x-rays (EDAX). Well oriented crystals were cut appropriately by an electric discharge cutting machine for direction dependent measurements. The magnetization data were measured in a Quantum Design Superconducting Quantum Interference Device (SQUID) magnetometer and Vibration Sample Magnetometer (VSM) in the temperature range 1.8 to 300~K and fields up to 14~T. The electrical resistivity, magnetoresistivity and the heat capacity were measured  in a Quantum Design Physical Properties Measurement System (PPMS). Heat capacity measurements down to 100~mK were performed using the dilution insert of QD-PPMS. A piston-cylinder type pressure cell  fabricated from MP35N alloy was used to measure the resistivity upto 12.6~kbar. A teflon capsule covering sample platform, holding two samples and one Sn wire as manometer, was  filled with Daphne oil, which is hydrostatic pressure medium. The dimension of capsule is such that it fits exactly inside the 5~mm bore of the cell. A feed through made out of beryllium copper alloy was used for providing electrical contacts to the sample and Sn manometer. A cernox temperature sensor and a heater in the form of manganin wire mounted over the cell were fed to a Cryocon PID temperature controller. A 5~mA dc current derived from Keithley current source was passed through samples and the generated voltage was read out by a sensitive nanovoltmeter. All the resistivity measurements were done in the standard four probe configuration. Temperature controller, current source and voltmeter were interfaced to a computer via GPIB and data acquisition was automated by a Labview programme.

\section{Results and Discussion}
\subsection{Structure}
The compositions obtained from EDAX analysis confirmed the stoichiometric ratio of 1:1:4:2 to within 1 at.\% for each element. The powder x-ray diffraction patterns of CeTAl$_4$Si$_2$ (T~=~Rh and Ir) are similar to those of Eu-analogs, and no extra peaks due to any parasitic phases were found.
%
%****************************Table1******************************
\begin{table}[h]
\caption{\label{TLP} Lattice constants $a$ and $c$, and unit cell volume $V$ of CeRhAl$_4$Si$_2$ and CeIrAl$_4$Si$_2$ as determined from the x-ray powder diffraction pattern.}
\begin{ruledtabular}
\begin{tabular}{lccc}
%\hline\hline
                 & $a$ & $c$ & $V$  \\
                 & (\AA)     & (\AA)     & (\AA$^3$)  \\
\hline
CeRhAl$_4$Si$_2$ & 4.223(2) & 8.048(2)   & 143.54(2)   \\ 
CeIrAl$_4$Si$_2$ & 4.236(3)  & 8.043(2)  & 144.34(2)   \\       
% \hline       
\end{tabular}
\end{ruledtabular}
\end{table}
%*****************************************************************
%
A Rietveld refinement using FullProf software~\cite{Carvajal} based on the EuIrAl$_4$Si$_2$-type tetragonal crystal structure was performed. The obtained lattice parameters $a$ and $c$  are listed in Table~\ref{TLP} and are in good agreement with the values reported in ref.~\onlinecite{Ghimire}. It may be noted that similar to Eu compounds~\cite{Maurya} the lattice parameter $a$ is larger but $c$ is slightly shorter in Ir analog compared to that of Rh analog but overall the unit cell volume of CeIrAl$_4$Si$_2$ is slightly larger than that of the Rh-analog, which is in accordance with the larger atomic volume of Ir. 
\subsection{Magnetic susceptibility and magnetization}
\begin{figure*}
\includegraphics[width=0.85\textwidth]{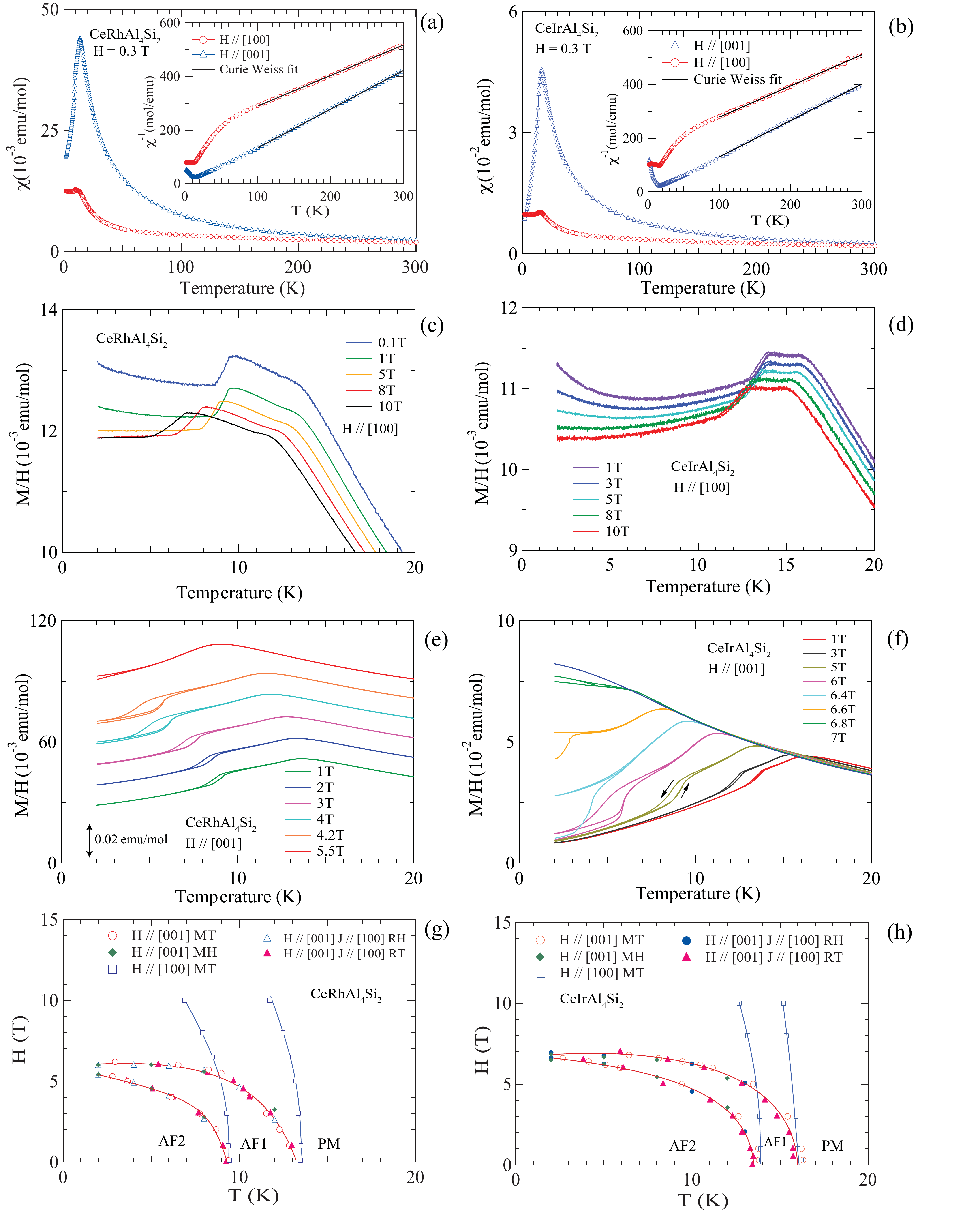}
\caption{\label{MT1}(Color online) Magnetic susceptibility and inverse susceptibility (in inset) up to 300~K of (a) CeRhAl$_4$Si$_2$ and (b) CeIrAl$_4$Si$_2$. The dependence of $M/H$ on field is shown in (c-f) for $H~\parallel$~[100] and $H~\parallel$~[001], respectively. In (e) and (f) single color plots show  ZFC (zero field cooled), FCC (field cooled cooling) and FCH (field cooled heating) data revealing field induced first order nature of transition at $T_{\rm N2}$. Magnetic phase diagrams of (g) CeRhAl$_4$Si$_2$ and (h) CeIrAl$_4$Si$_2$ derived from magnetization (MT, MH) and magnetoresistivity (RT, RH) data (\textit{vide infra}). AF1, AF2 and PM represents two antiferromagnetic and paramagnetic phases, respectively.}
\end{figure*}

The magnetic susceptibility, $\chi(T)$, of CeRhAl$_4$Si$_2$ and CeIrAl$_4$Si$_2$ below 300~K is shown in the main panel of Figs.~\ref{MT1} (a) and~\ref{MT1} (b), for field (0.3~T) applied along the [100] and [001] directions, respectively. The inverse susceptibility in the temperature range 1.8 to 300~K is plotted in the insets of Fig.~\ref{MT1}(a) and (b). The susceptibility is highly anisotropic in the paramagnetic region in both compounds and a fit of the Curie-Weiss law, $\chi~= \frac{C}{T - \theta_{\rm p}}$ to the high temperature data (100-300~K), represented by the solid lines,  furnishes the Curie-Weiss parameters  which are listed in Table~\ref{CW_parameters}.
%
%*********************FIGURE 1 **********************************
%****************************************************************
%
 The effective moments are close to the Ce$^{3+}$-free ion moment value of 2.54~$\mu_{\rm B}$/Ce. The highly anisotropic nature of the susceptibility is clearly reflected by the respective values of paramagnetic Curie temperature $\theta_{\rm p}$ which are more than one order of magnitude larger for $H~\parallel$~[100] in both compounds.  In conformity with the trivalent nature of the Ce ions, the compounds order antiferromagnetically at $T_{\rm N1}$~=~13.3 and 16~K, and $T_{\rm N2}$~=~9.4 and 13.8~K for T~=~Rh and Ir, respectively, close to the values reported in ref.~\onlinecite{Ghimire}. It may be noted that the absolute value of the polycrystalline averaged $\theta_{\rm p}$ ($-100$ and $-68$~K) is substantially higher than $T_{\rm N1}$. We believe it to be primarily due to the crystal electric field and anisotropic Kondo interaction (\textit{vide infra}) which contribute negatively to $\theta_{\rm p}$. In the simplest collinear two-sublattice antiferromagnet, the $\chi_{\perp}$ along the hard direction is temperature independent below $T_{\rm N}$ while $\chi_{\parallel}$ gradually decreases to zero as T$\rightarrow$0. A weak temperature dependence of $\chi$ along [100] in the two compounds indicates a hard $ab$-plane. 
 
 It may be noted that the anisotropy in $\chi_[100]$ and $\chi_[001]$ persists in our data(Figs.~\ref{MT1}a and ~\ref{MT1}b)upto 300~K. On the other hand in ref.~\onlinecite{Ghimire} the two cross each other near 275~K in CeIrAl$_4$Si$_2$. Such a crossover, prima-facie indicates a change in the easy axis of magnetization with temperature. However, in the crystal electric field analysis of the magnetization data (vide infra), the values of the dominant CEF parameter B$_2^0$ are negative in both compounds and comparable making the anisotropic crossover along the crystallographic axis unlikely. 
%****************************Table2******************************
\begin{table}[h]
\caption{\label{CW_parameters} Effective moment and paramagnetic Curie temperature in CeRhAl$_4$Si$_2$ and CeIrAl$_4$Si$_2$ along the principal crystallographic directions.}
\begin{ruledtabular}
\begin{tabular}{ccccc}
%\hline \hline
 &\multicolumn{2}{c}{CeRhAl$_4$Si$_2$} & \multicolumn{2}{c}{CeIrAl$_4$Si$_2$} \\ \hline
 & $\mu_{\rm eff}(\mu_{\rm B}/f.u.)$ & $\theta_{\rm p}\rm(K)$ & $\mu_{\rm eff}(\mu_{\rm B}/f.u.)$ & $\theta_{\rm p}\rm(K)$   \\ \hline
$H~\parallel$~[100] & $2.65$ & $-155$ & $2.62$ & $-140$ \\
$H~\parallel$~[001] & $2.35$ & $7.1$ & $2.43$ & $4$ \\
%\hline
\end{tabular}
\end{ruledtabular}
\end{table}
%****************************************************************
%

The magnetic field dependence of susceptibility ($M/H$) below $T_{\rm N1}$ was investigated at a few fields and the data are plotted in Figs.~\ref{MT1}(c-f). The $T_{\rm N}$ decreases as the applied field is increased, as commonly observed in antiferromagnets. However, the decrease is more substantial for $H~\parallel$~[001] which is relatively the easy axis of magnetization. The magnetization in both compounds at $T_{\rm N2}$ exhibits hysteresis thereby indicating that the transition at $T_{\rm N2}$ has a first order character (Figs. 1e-f). The hysteresis is weaker at lower fields and it increases with increment in magnetic field till spin flop field. The change in $T_{\rm N1}$ and $T_{\rm N2}$ with field applied along [100] and [001] directions has been summarized in the phase diagrams shown in Figs.~\ref{MT1}g and \ref{MT1}h. Critical points obtained from isothermal magnetization $vs.$ magnetic field (MH), electrical resistivity as a function of field (RH) and temperature (RT) measurements are also shown which are in consonance with each other. It may be noted that the phase boundary separating two antiferromagnetic  states AF1 and AF2 for H~$\parallel$~[001] is of first order character as inferred from the susceptibility and magnetoresistivity data. Also, $T_{\rm N}$s suppress faster for field applied along the easy axis \textit{viz.} [001].

%*********************FIGURE 2**********************************
\begin{figure*}
\includegraphics[width=0.85\textwidth]{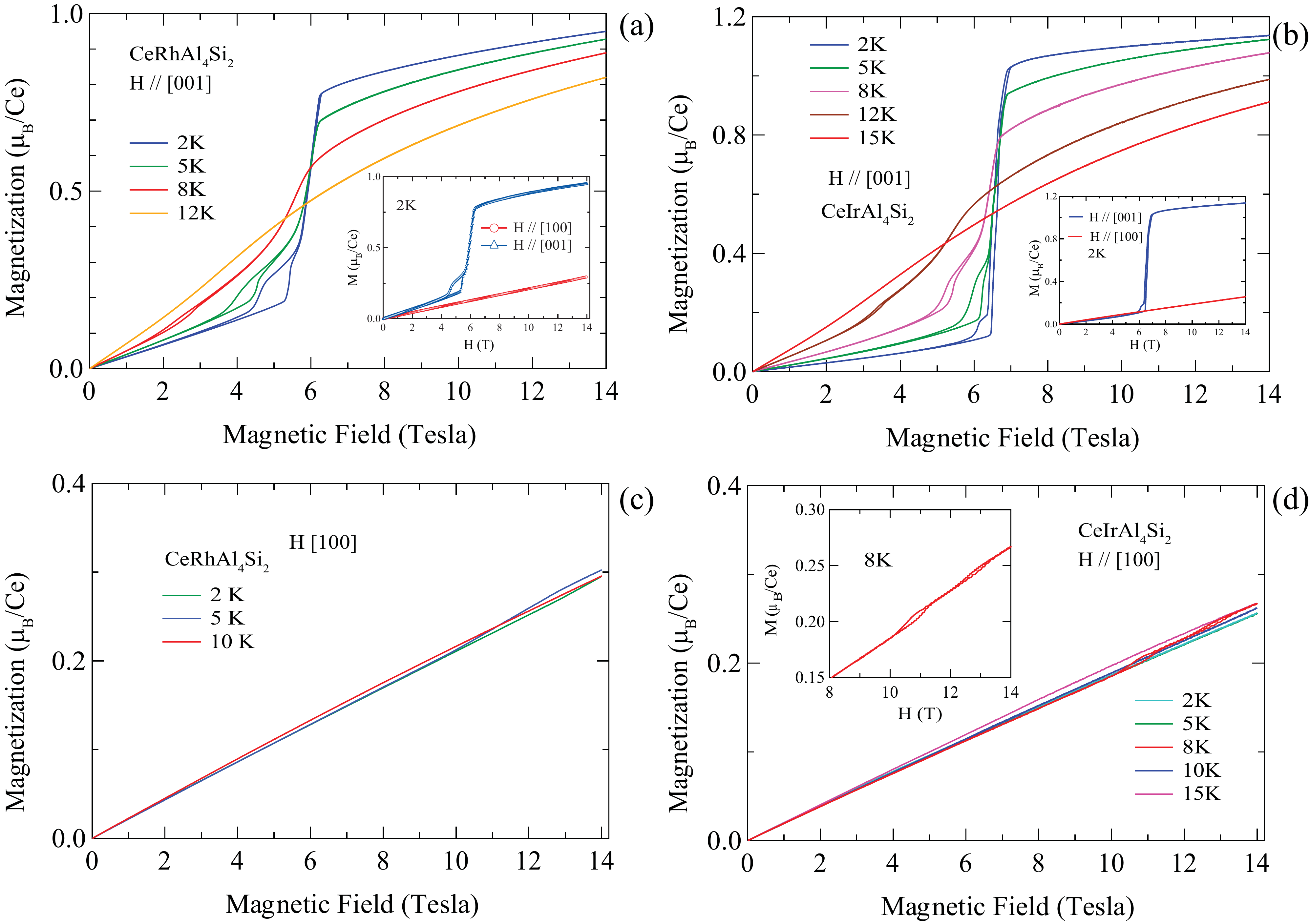}
\caption{\label{MH1}(Color online) Isothermal magnetization curves at selected temperatures for (a) CeRhAl$_4$Si$_2$ and (b) CeIrAl$_4$Si$_2$ along [001] direction. The insets show the data at 2~K for H~$\parallel$~[001] and H~$\parallel$~[100]. (c) and (d) show the data at selected temperatures for H~$\parallel$~[100]. The inset in (d) reveals two weakly first order changes which are also seen in the magnetoresitance data.(see Fig.~\ref{MR2}).}
\end{figure*}
%****************************************************************
%
The isothermal magnetization at 2~K is linear up to 14~T along [100] while there is a sharp spin-flop (metamagnetic-like) transition revealed by two closely spaced jumps in the magnetization along the [001]-direction beginning at 5.3(6.4)~T for CeRh(Ir)Al$_4$Si$_2$ (see Figs.~\ref{MH1}a and~\ref{MH1}b). The former behaviour is typical of an antiferromagnet when the moments are perpendicular to the field. The field dependence thus clearly marks the easy and the hard-axes of
 magnetization.  This is in conformity with neutron diffraction in which  magnetic moments aligned antiferromagnetically along the \textbf{c}\--axis below $T_{\rm N2}$ is inferred~\cite{Ghimire_Neutron}. The magnetization attains a value of 0.95 and 1.14~$\mu_{\rm B}$/Ce at 14~T along
  the \textbf{c}\--axis in CeRhAl$_4$Si$_2$ and CeIrAl$_4$Si$_2$ respectively, which is lower than the saturation moment of Ce$^{3+}$
   (2.14~$\mu_{\rm B}$/Ce). We attribute the lower values to the combined effects of crystal electric field and partial
  quenching of the Ce moments due to Kondo screening. We note that the magnetization at 14~T at 2~K is lower in the Rh-analogue, which is consistent with its higher Kondo temperature, $T_{\rm K}$ (\textit{vide infra}). The magnetization at 2~K in 14~T is not yet saturated and is lower than the values determined in ref.~\onlinecite{Ghimire_Neutron}.   
%
%*********************FIGURE 3 **********************************
\begin{figure*}
\includegraphics[width=0.85\textwidth]{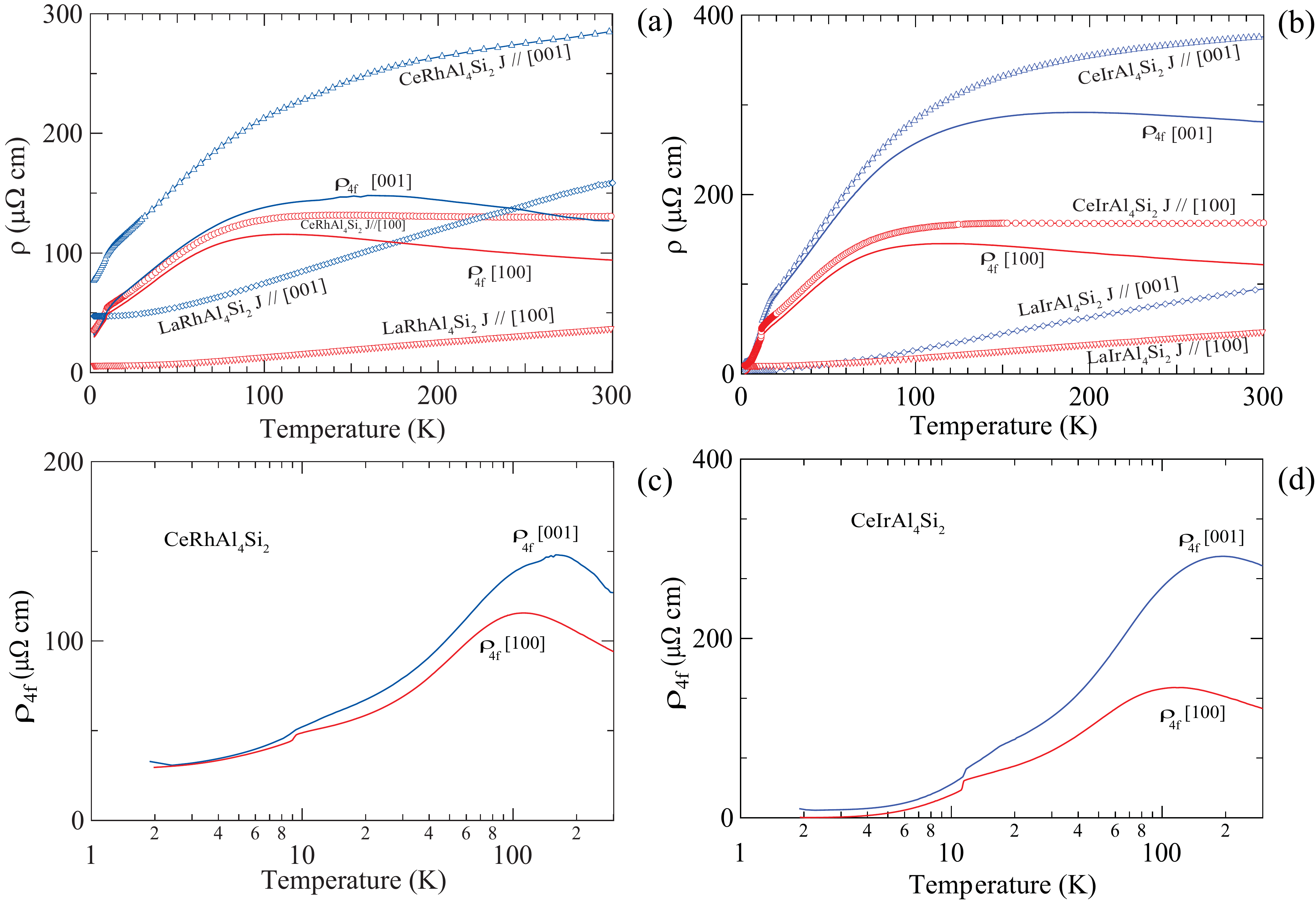}
\caption{\label{RT1}(Color online) Electrical resistivity of (a) CeRhAl$_4$Si$_2$ and LaRhAl$_4$Si$_2$; (b) CeIrAl$_4$Si$_2$ and LaIrAl$_4$Si$_2$ along the major crystallographic directions. $\rho_{\rm 4f}$ is represented by solid lines. $\rho_{\rm 4f}(T)$ data plotted on logarithmic temperature scale are shown in (e) and (f).}  
\end{figure*}
%****************************************************************

	One can observe a clear hysteresis in the vicinity of spin-flop transition. The hysteretic behaviour is also observed in the temperature dependence of magnetic susceptibility (mentioned above) and electrical resistivity (\textit{vide infra}), revealing the presence of first order field induced effect. The spin-flop field value and the magnetization decrease with the increase of temperature. The magnetization for $H~\parallel$~[100] is relatively insensitive to the variation in temperature as inferred from Figs.~\ref{MH1}(c) and~\ref{MH1}(d). The inset in Fig.~\ref{MH1}(d) reveals two weakly hysteretic regions in the magnetization of CeIrAl$_4$Si$_2$ at 8~K. 
\subsection{Electrical Resistivity}
Figs.~\ref{RT1}(a-d) show the zero-field electrical resistivity $\rho (T)$ data of CeRhAl$_4$Si$_2$ (left panels) and CeIrAl$_4$Si$_2$ (right panels) for the current density $J$ parallel to [001] and [100] directions, respectively. The corresponding data for the non-magnetic La-reference compounds are also plotted. While anomalies at $T_{\rm N1}$ and $T_{\rm N2}$ for $J~\parallel$~[001] are visible either in the $\rho$ vs $T$ or $d\rho /dT$ vs $T$ plots (not shown), the $\rho$ for $J~\parallel$~[100] shows a sudden change of slope only at $T_{\rm N2}$. 

There is a considerable anisotropy in the resistivity, $\rho$ along [001] being larger compared to [100] in the entire temperature range. The residual resistivity ratio (\textit{RRR}) $\rho_{300K}/\rho_{2K}$ is 17.8 and 29.9 for $J~\parallel$~[100] and [001], respectively for CeIrAl$_4$Si$_2$, compared to its corresponding values of 3.2 and 3.4 in CeRhAl$_4$Si$_2$. While our \textit{RRR} for CeIrAl$_4$Si$_2$ is comparable to the value reported by Ghimire et al~\cite{Ghimire}, in CeRhAl$_4$Si$_2$ our \textit{RRR} values are lower for reasons unknown to us. We repeated the measurements on a second sample of CeRhAl$_4$Si$_2$ but observed similar low values of \textit{RRR}. Our values of resistivity are also significantly higher  than reported in ref.~\onlinecite{Ghimire}. Though our single crystals looked good superficially, there may be micro-voids and micro-cracks which lead to higher observed resistivity.  While $\rho_{[001]}$ decreases as the temperature is decreased below 300~K, $\rho_{[100]}$ shows a slight negative temperature coefficient above 100~K. The $4f$-derived part of the resistivity $\rho_{\rm 4f}$ is calculated by subtracting the $\rho (T)$ data of La-analog from the corresponding Ce-compound, which is also shown in Figs.~\ref{RT1}(a) and~\ref{RT1}(b) and replotted in~\ref{RT1}(c) and~\ref{RT1}(d) on a semi-logarithmic scale. $\rho_{\rm 4f}$ reveals a negative logarithmic temperature dependence along both directions which is a hallmark of the Kondo interaction. The high temperature peak in $\rho_{\rm 4f}$ in range 100-200~K, which arises due to the interplay of Kondo interaction and crystal electric field levels, occurs at different temperatures along the two directions. The resistivity data thus reveal that these two Ce-compounds are dense anisotropic Kondo lattice antiferromagnets.

%*********************FIGURE 4**********************************
\begin{figure*}[!]
\includegraphics[width=0.95\textwidth]{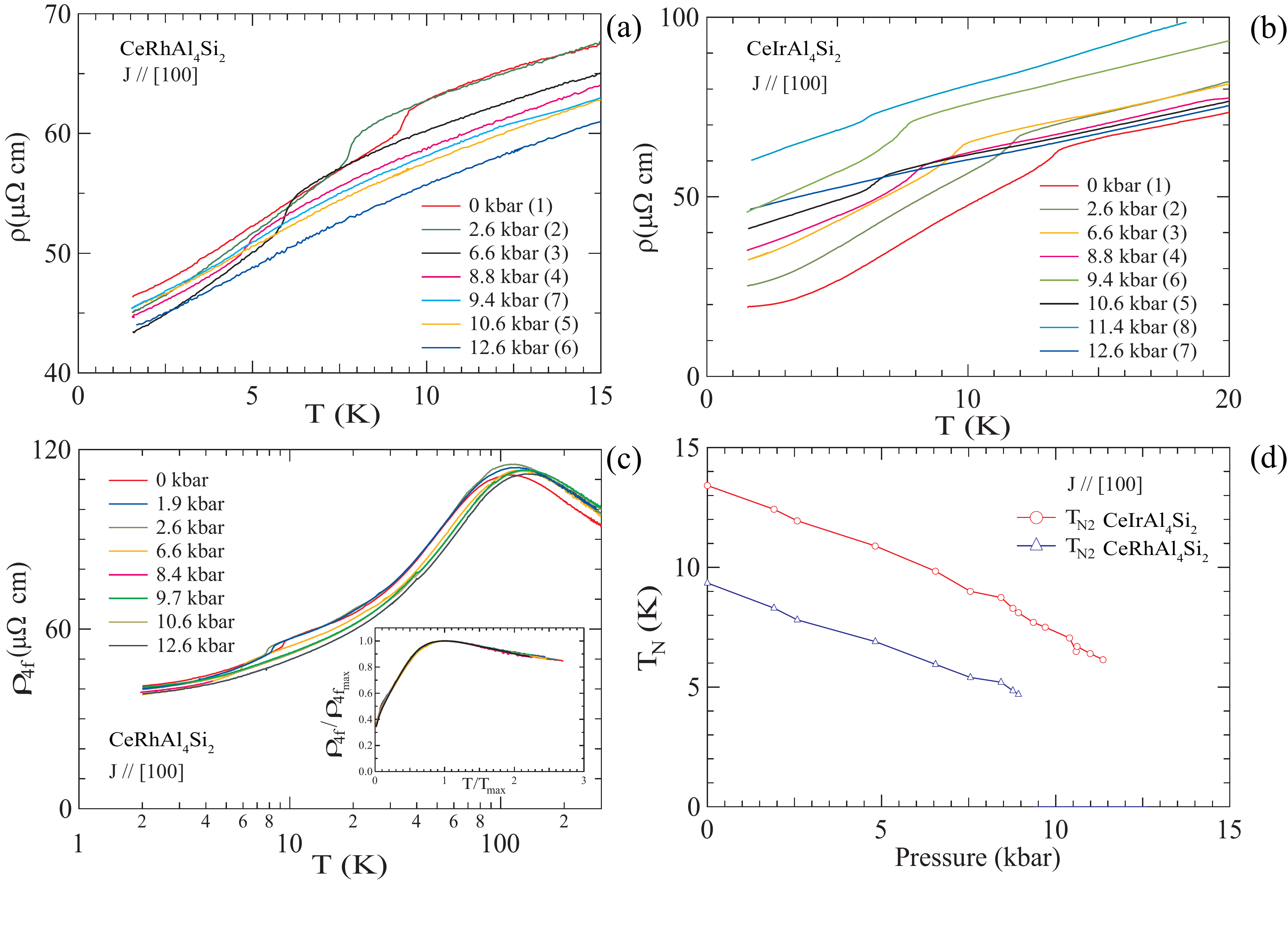}
\caption{\label{Pressure}(Color online) Electrical resistivity under hydrostatic pressure at low temperatures when $J~\parallel$~[100] for (a) CeRhAl$_4$Si$_2$, (b) CeIrAl$_4$Si$_2$. Numbers in parenthesis denote the the order in which pressure was applied. (c) $4f$\--derived electrical resistivity of CeRhAl$_4$Si$_2$ upto 300~K on logarithmic temperature scale, (d) variation of $T_{\rm N2}$ with pressure in CeRhAl$_4$Si$_2$ and CeIrAl$_4$Si$_2$ when $J~\parallel$~[100].}
\end{figure*}
%****************************************************************	

The Kondo behaviour of CeTAl$_4$Si$_2$ compounds, inferred above from the resistivity data at ambient pressure,  was probed further by measuring the resistivity under pressure up to 12.6~kbar. According to the standard Doniach phase diagram of a magnetic Kondo lattice, pressure enhances the $4f$\--conduction band coupling and the Kondo temperature, which decreases the $T_{\rm N}$. At sufficiently high pressures when the Kondo interaction dominates over the RKKY interaction, $T_{\rm N}$ approaches zero leading to a quantum phase transition. Fig.~\ref{Pressure} shows the resistivity of the two compounds at selected values of pressure. The pressure cell was simultaneously loaded with the two compounds and numbers in  parentheses indicate the order of data acquisition.  The $T_{\rm N2}$ in both compounds decreases with pressure, which is qualitatively in consonance with the Doniach phase diagram.  $T_{\rm N2}$ decreases from 13.6 (at ambient pressure) to 8.2 (6.2)~K at 8.8(11.4)~kbar in CeIrAl$_4$Si$_2$, while it decreases from 9.4 to 4.7~K at 8.8~kbar in CeRhAl$_4$Si$_2$. No apparent anomaly is observed at higher pressures in the two compounds down to 1.8~K both in $\rho$ versus $T$ and $d\rho/dT$ versus $T$ plots.  Data at lower temperatures and presumably higher pressures are required to track the eventual decrease of $T_{\rm N2}$ to 0~K. 

	The resistivity data under pressure were measured up to 300~K and the data for CeRhAl$_4$Si$_2$ are shown in Fig.~\ref{Pressure}c. The temperature $T_{max}$ at which the resistivity attains its maximum value $\rho _{max}$ increases from $\sim$~108 at ambient pressure to 130~K at 12.6~kbar. Qualitatively, the upward shift of $T_{max}$ is in conformity with the increase of $T_{\rm K}$ with pressure. The inset of Fig.~\ref{Pressure}c shows $\rho / \rho _{max}$ scales with $T / T_{max}$ in a reasonably large interval around $T_{max}$. Such a scaling relationship has been reported in some Kondo compounds~\cite{Thompson}. Though $T_{max}$ increases with pressure in the Ir-analogue as well, we observed a substantial difference in the resistivity values particularly at higher temperatures taken at nearly the same pressure in two different runs (without affecting the transition temperature), while as in the Rh compound the resistivity values nearly matched. Therefore, we have not shown the data for the Ir-anologue at higher temperatures. Fig.~\ref{Pressure}d shows the T-P phase diagram of both compounds. It may be noted that $T_{\rm N1}$ is not discernible in any of our pressure dependent resistivity measurement as the data were taken for $J~\parallel$~[100] where we see only one transition (\textit{cf.} Figs.~\ref{RT2}c and ~\ref{RT2}d). 
	 
\subsection{Magnetoresistance}

\begin{figure*}
\includegraphics[width=0.85\textwidth]{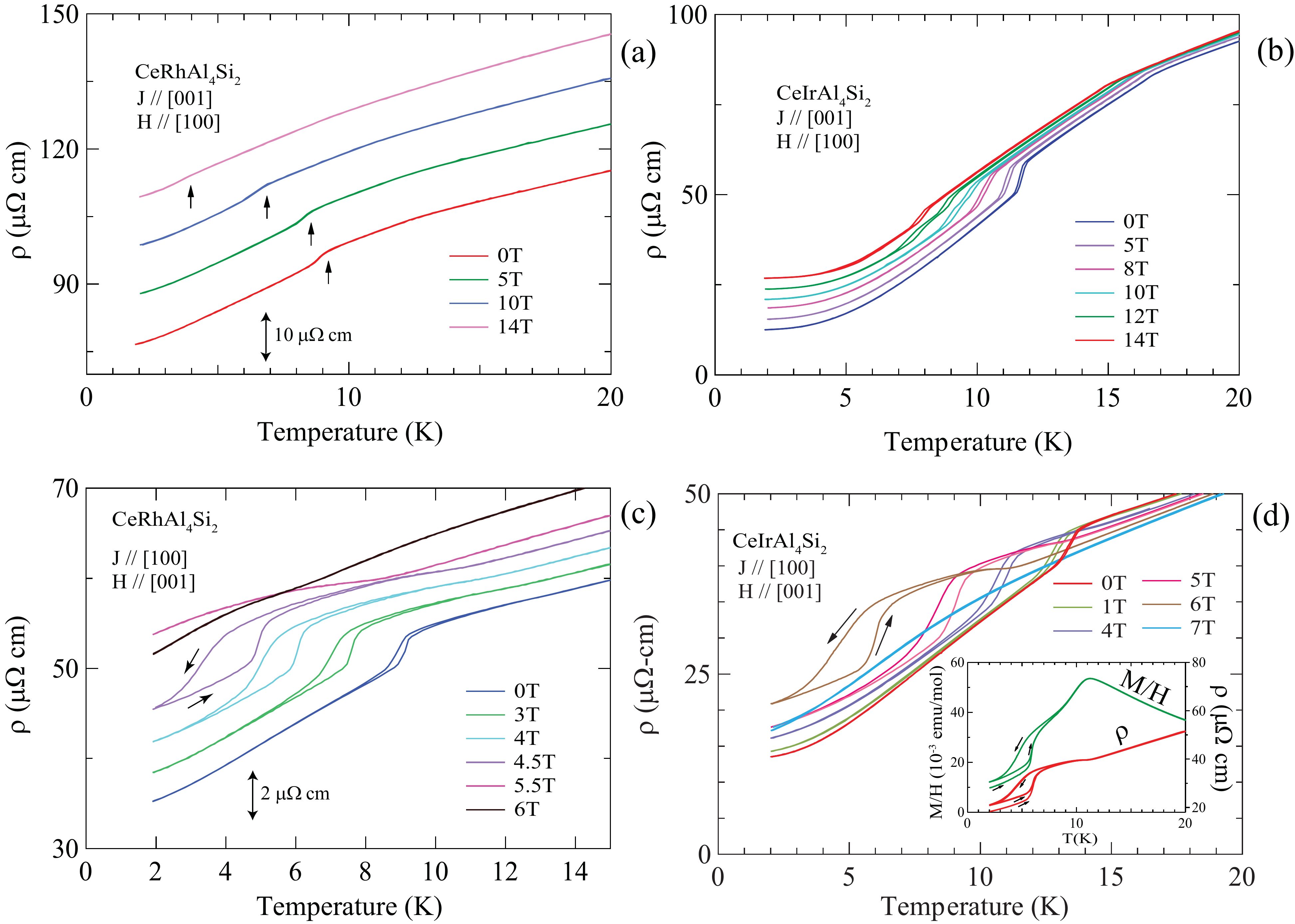}
\caption{\label{RT2}(Color online) Magnetic field dependence of $\rho (T)$ of CeRhAl$_4$Si$_2$ and CeIrAl$_4$Si$_2$ in different configuration.  In (a) and (c) the in-field plots have been shifted upward by a constant offset for clarity.}
\end{figure*}

The variation of resistivity at selected values of magnetic field at low temperatures is shown in Figs.~\ref{RT2}(a-d). For $H$ applied along [100] and $J~\parallel$~[001] the resistivity plots are qualitatively similar except that there is a gradual downward shift of $T_{\rm N}$ with increasing field. In CeIrAl$_4$Si$_2$, one sees two steps to the lower temperature transition at fields of 10 and 12~T (Fig.~\ref{RT2}b), which is consistent with two weakly first order changes seen in the magnetization (inset, Fig.~\ref{MH1}d). On the other hand prominent changes are seen for $H~\parallel$~[001] and $J~\parallel$~[100]. $T_{\rm N}$ is suppressed relatively faster with a substantial hysteresis in the field\--range where spin\--flop occurs, revealing first order effects in conformity with the magnetization data. It may be noted that there is a slight kink at 8.6 and 11.2~K in 5.5~T and 6~T in CeRhAl$_4$Si$_2$ and CeIrAl$_4$Si$_2$, respectively. The inset of Fig.~\ref{RT2}(d) shows that $T_{\rm N1}$ in 6~T has decreased to the temperature at which the kink occurs in CeIrAl$_4$Si$_2$. Maybe the anomaly is present at lower fields also but it is not discernible. Similar explanation holds in case of CeRhAl$_4$Si$_2$. The inset also shows that there is a good agreement between the hysteresis in the resistivity  (on right scale) and $M/H$ (on left scale) data at  6~T field applied parallel to [001]. 

%*********************FIGURE 5***********************************
%****************************************************************

%*********************FIGURE 6 **********************************
\begin{figure*}
\includegraphics[width=0.85\textwidth]{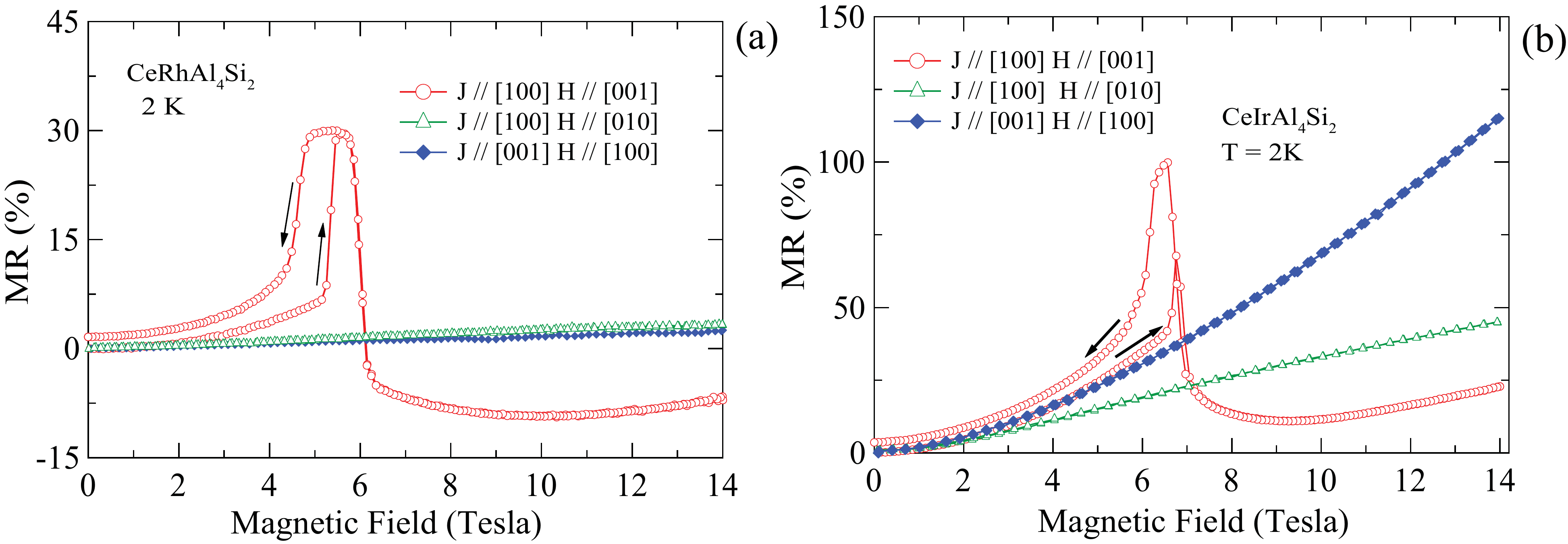}
\caption{\label{MR1}(Color online) Current and field-direction dependence of isothermal Magnetoresistance($MR$) at 2~K upto 14~T of (a) CeRhAl$_4$Si$_2$ and (b) CeIrAl$_4$Si$_2$.}
\end{figure*}
%****************************************************************

\begin{figure*}
\includegraphics[width=0.75\textwidth]{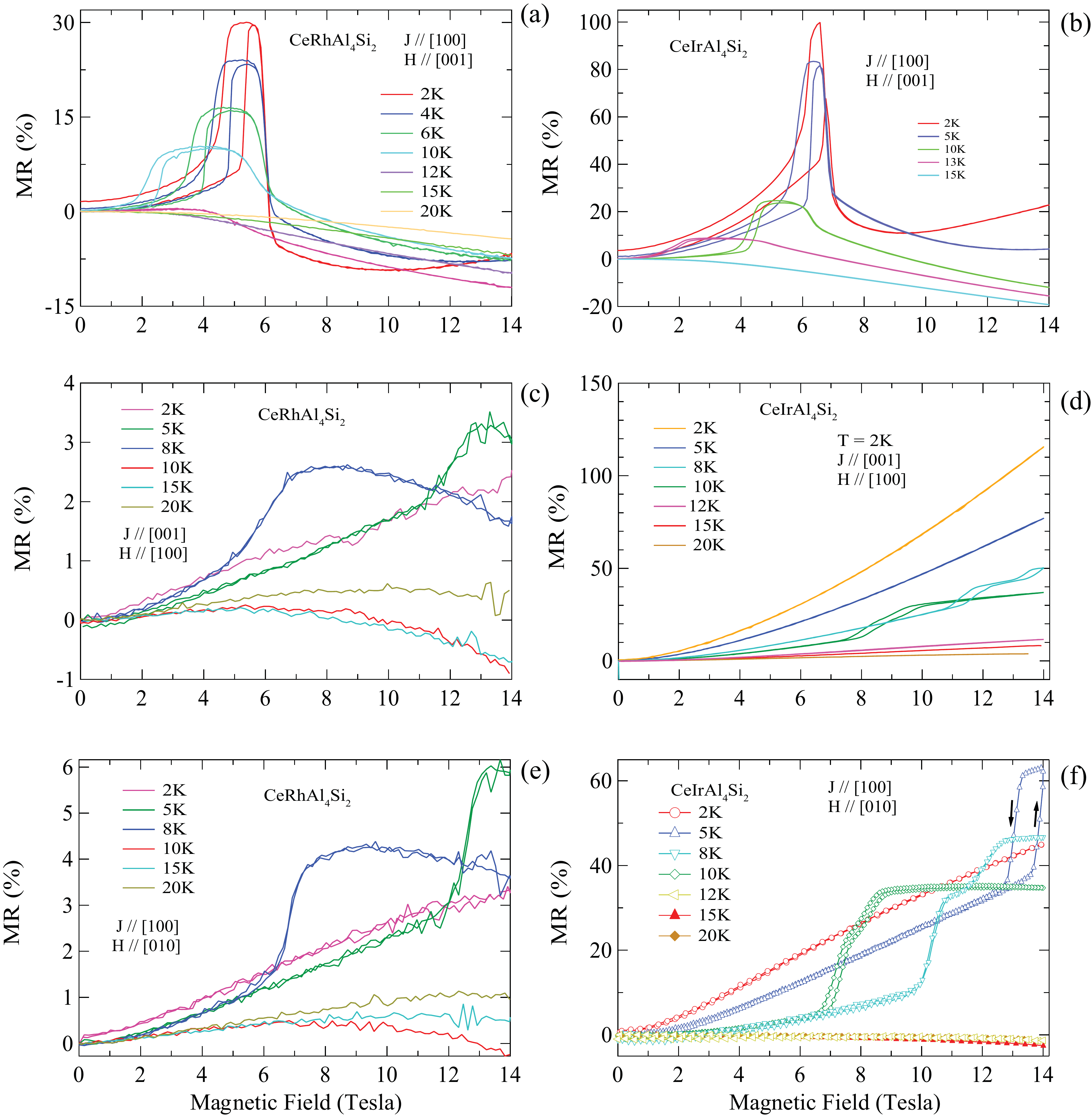}
\caption{\label{MR2}(Color online) Temperature dependence of $MR(H)$ in  CeRhAl$_4$Si$_2$ and CeIrAl$_4$Si$_2$. The data in (c) and (e) are looking scattered due to the relatively smaller values of the $MR$.}
\end{figure*}

	Magnetoresistivity $MR$ was also probed by varying the field from zero to 14~T at selected temperatures, with the field and current density in various orientations and the results are plotted in Figs.~\ref{MR1} and~\ref{MR2}. The magnetoresistance, $MR$ is defined as $MR=[\rho(H)-\rho(0)]/ \rho(0)$. With reference to Fig.~\ref{MR1}(a) the $MR$ of CeRhAl$_4$Si$_2$ at 2~K for field applied in the $ab$-hard plane is small and attains a value of $\sim$~3 \% at 14~T. On the other hand for field applied along the easy-axis [001] the $MR$ at its maximum is an order of magnitude larger. Initially the positive $MR$ increases with field, jumps sharply at the first step of the spin-flop attaining a maximum of ~30\% at nearly the second step and then decreases sharply and becomes negative at higher fields. In the return cycle the $MR$ shows hysteresis in the spin-flop region, which corresponds nicely with the hysteresis in the magnetization (\textit{cf.} Fig.~\ref{MH1}).  We have reported earlier simialr sharp changes in the $MR$ of EuNiGe$_3$ which is an antiferromagnet with $T_{\rm N}$ of 13.2~K~\cite{Maurya_EuNiGe3}.

%*********************FIGURE 7*********************************
%****************************************************************	
	
	For CeIrAl$_4$Si$_2$ the $MR$ for $H~\parallel$~[001] is qualitatively similar (Figs.~\ref{MR1} (b) at low fields except that the $MR$ in the spin-flop region is even larger and exceeds 100 \%. Above the spin-flop the $MR$ drops sharply but unlike CeRhAl$_4$Si$_2$ it does not attain negative values and remains positive up to the highest field of 14~T. For field applied in the $ab$-hard plane viz. $J~\parallel$~[100] and $H~\parallel$~[010]; $J~\parallel$~[001] and $H~\parallel$~[100] the $MR$ is positive qualitatively similar the that of CeRhAl$_4$Si$_2$ but its magnitude is much larger. At 2~K the $MR$ rises monotonically reaching 44 and 115~\%, respectively at 14~Tesla. The latter value is even larger than the peak value for $H~\parallel$~[001].
The observed $MR$ can be qualitatively explained by invoking three contributions, i) the $MR$ of an antiferromagnet, ii) the Kondo contribution and iii) the positive contribution due to the cyclotron motion of the conduction electrons. In an antiferromagnet for $T << T_N$ when the field is applied along the easy axis, it tends to suppress the spin fluctuations in one sublattice and increase in the other. Typically, the $MR$ resulting from the combined fluctuations is positive~\cite{Fournier}. In the spin-flop region the spins can be imagined to be in a canted state with higher spin disorder resistivity. As the field is increased further leading to spin-flip where a saturated paramagnetic state is achieved, the $MR$ becomes negative. 
A magnetic field tends to suppress  the Kondo scattering of the conduction electrons leading to a negative $MR$.  According to Zlatic's model, which is based on the single center scattering picture, the $MR$ has a negative minimum at nearly $T=T_{\rm K}/2$, changes sign at around T$_K$/2$\pi$ and has a positive maximum at $T=0~K$~\cite{Zlatic}. Taking a $T_{\rm K}$ of nearly 10~K for CeRhAl$_4$Si$_2$ (\textit{vide infra}) the negative minimum due to Kondo interaction should occur around 5~K. In CeIrAl$_4$Si$_2$ with a lower $T_{\rm K}$, the negative minimum should occur at even lower temperature. The $MR$ is positive at these temperatures indicating that the contribution due to the effect of magnetic field on the antiferromagnetic state to $MR$ is dominant. The negative $MR$ of CeRhAl$_4$Si$_2$ above 6~T may have a contribution from the Kondo
 interaction while it would be relatively far lower in CeIrAl$_4$Si$_2$. As regards the positive contribution from the cyclotron motion of the electrons, in a simple two-band model~\cite{Jan} the $MR$ is proportional to $B^2$ where $B$ is the field and inversely proportional to $\rho (0)^2$.
  The resistivities of CeRhAl$_4$Si$_2$ and CeIrAl$_4$Si$_2$ at 2~K are 35.26 and 13.53 $\mu\Omega cm$ for $J~\parallel$~[100];
  the ratio decreases with increasing temperature. Therefore, it is expected that the positive cyclotron $MR$ at 2~K will be roughly nine times larger in CeIrAl$_4$Si$_2$. We speculate that may be the reason why the $MR$ remains positive in the Ir-analogue at high fields while it is negative in the Rh-analogue.

	The sharp upturn in the $MR$ at spin flop and the associated hysteresis shift to lower fields as the temperature is increased (Figs.~\ref{MR2} (a) and~\ref{MR2} (b)). The relatively sharp peak in the $MR$ at 2~K increasingly broadens at higher temperatures, mimicking qualitatively the broadened spin-flop region as the temperature is increased (cf Figs. 2a and 2b). Above the spin flop, qualitatively the $MR$ shows a similar field and temperature dependence in two compounds. As mentioned above the Kondo scattering of the conduction electrons should be partially quenched in a magnetic field leading to negative MR. The effect is maximum close to T$_K$/2 and decreases as the temperature is increased. It may be noted that at 4, 6 and 10~K the $MR$ in CeRhAl$_4$Si$_2$ in a limited range above 6~T becomes less negative than its corresponding values at 2~K. A negative contribution from the Kondo interaction that decreases as the temperature is increased above T$_K$ may explain this. As the Kondo interaction is relatively weaker in the Ir-analogue, its contribution to $MR$ is not significant enough to cause a similar behaviour in   CeIrAl$_4$Si$_2$. Possible differences in the positive $MR$ due to the cyclotron motion of the conduction electrons between the two compounds decrease with the increase of temperature.  In the paramagnetic region the applied field will tend to suppress residual spin fluctuations leading to a negative $MR$ as observed at 15 and 20~K in CeRhAl$_4$Si$_2$. Clearly several competing mechanisms are contributing to $MR$ which have slightly different comparative energy scales in the two compounds.  
		
	The $MR$ in CeRhAl$_4$Si$_2$ at selected temperatures, for fields applied in the $ab$-plane, i.e. along [100] and [010] is qualitatively similar for current density parallel to [001] and [100], respectively (see, Figs. 7c and 7e). A relatively sharp upturn in $MR$ is seen at 5  and 8~K around $\sim$~12 and $\sim$~6~T, respectively, which is most likely arising from changes in the spin reorientation. In CeIrAl$_4$Si$_2$,for $J~\parallel$~[001] and $H~\parallel$~[100], we observe two steps with hysteresis at 8~K which correlate well with the data shown in the inset of (Fig.~\ref{MH1}(d). A single step with hysteresis is seen in the 10~K data; however no corresponding anomaly is clearly seen in the magnetization. The $MR$ in CeIrAl$_4$Si$_2$ for $J~\parallel$~[100] and $H~\parallel$~[010] also shows some peculiar features. A field induced hysteresis appears at 5~K around ~13~T, which splits into two at 8 and 10~K at lower fields but vanishes again at 12~K. It may be noted that the ratios of resistivities of the Rh and Ir compounds at 2, 10 and 20~K for J parallel to [001] are nearly 6, 2.4 and 1.2, respectively. As mentioned above the positive $MR$ due to the cyclotron motion is expected to be much larger in the Ir analogue, particularly at low temperatures.

\subsection{Heat Capacity}

\begin{figure*}[!]
\includegraphics[width=0.85\textwidth]{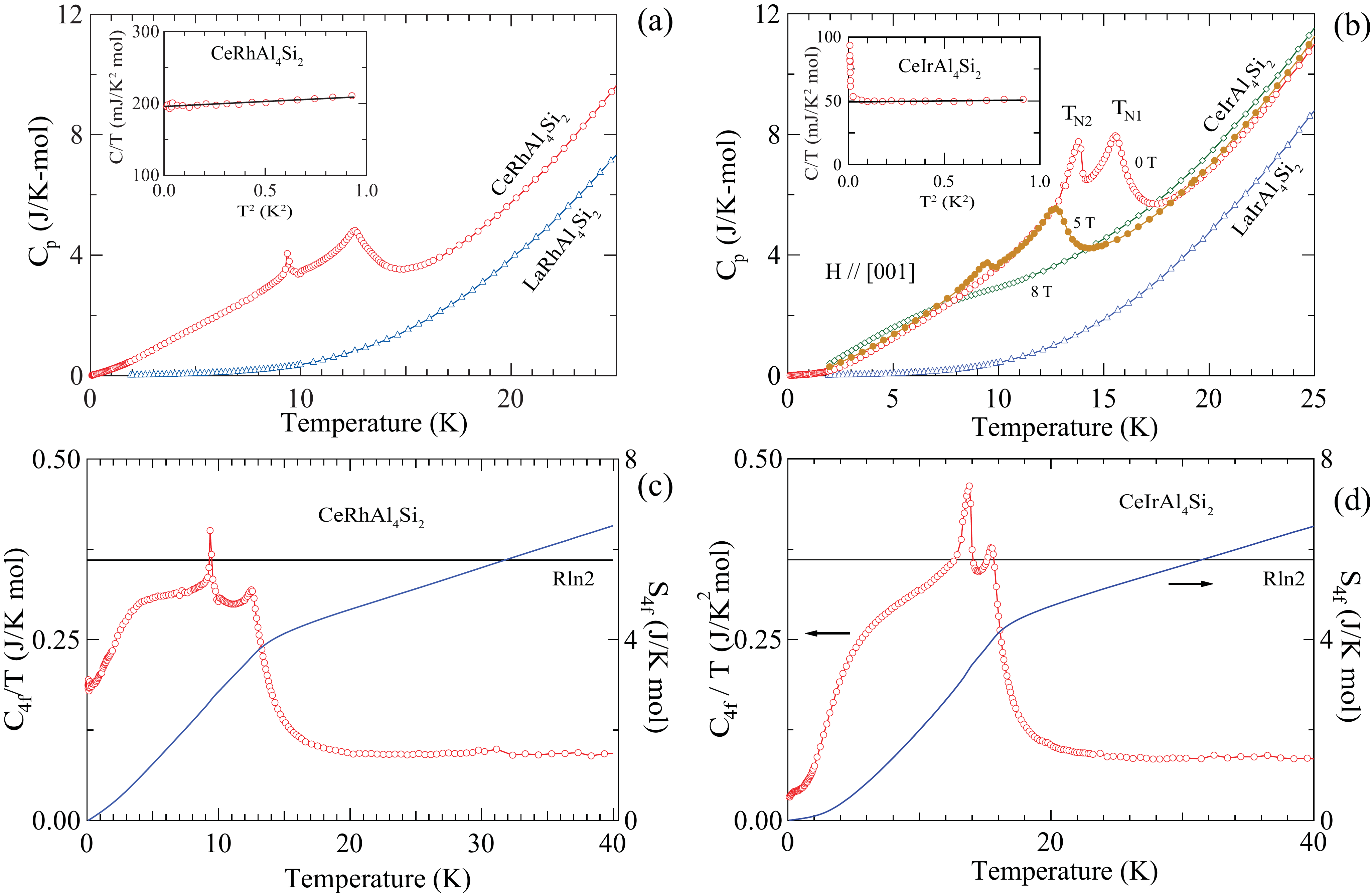}
\caption{\label{HC1}(Color online) Heat capacity and calculated entropy as a function of temperature of  CeRhAl$_4$Si$_2$ (left panels) and CeIrAl$_4$Si$_2$(right panels). The heat capacity of La analogs are also plotted. The inset in (a) and (b) show the low temperature data of Ce compounds as $C/T$ vs. $T^2$.}
\end{figure*}

 The heat capacity of CeRh(Ir)Al$_4$Si$_2$ was measured between 100~mK and 150~K to gain more information about the magnetically ordered Kondo lattice state such as the entropy associated with the magnetic ordering, values of the coefficient of the linear term in the electronic heat capacity  $\gamma$, which is proportional to enhanced  electron effective mass, and Kondo temperature. The heat capacity of  nonmagnetic analogs LaRh(Ir)Al$_4$Si$_2$ was also measured between 2 and 150~K. The data are plotted in Fig.~\ref{HC1}. The heat capacity clearly exhibits two peaks in both cerium compounds (Figs.~\ref{HC1} (a) and~\ref{HC1}(b)) confirming the occurrence of two bulk phase transitions. The peak temperatures are overall in good agreement with the corresponding peaks in the susceptibility. For CeIrAl$_4$Si$_2$ the heat capacity was also measured in 5 and 8~T.
 %*********************FIGURE 8**********************************
%****************************************************************  
   The two peaks shift to lower temperatures in 5~T and they virtually disappear in 8~T above the spin-flop field where the field induces a saturated paramagnetic state. 
   
   The heat capacity data of La analog is typical of a nonmagnetic reference compound. The plots of $C/T$ \textit{vs.} $T^2$ below 5~K for the two La-compounds (not shown) are linear and a fit of the standard expression $C/T~=~\gamma + \beta T^2$, where $\gamma$ and $\beta$ are the electronic and phononic part of the heat capacity, furnishes the following values, $\gamma$~=~8.7 and 8.0~mJ/mol K$^2$, $\beta$~=~0.22 and 0.21~mJ/mol K$^4$ for the Rh and Ir compounds, respectively. Our values of $\gamma$ and $\beta$ are comparable with those reported in ref.~\onlinecite{Ghimire}. The $4f$-derived entropy $S_{\rm 4f}$ was calculated by the following relation:
\begin{equation}
\label{entropy}
S_{\rm 4f} = \int \frac{C_{\rm 4f}}{T}dT
\end{equation}
where $C_{\rm 4f}$ was obtained by subtracting the heat capacity of the La-analog from the corresponding Ce-compound and making the usual assumption of lattice heat capacity being identical for the isotypic La and Ce compounds. Only 64\% (72\%) of entropy $S_{\rm 4f}$ for a doublet ground state with effective spin 1/2 (i.e. $Rln2$) is released up to $T_{\rm N1}$ for CeRh(Ir)Al$_4$Si$_2$ indicating the presence of Kondo interaction in these materials, assuming insignificant short range order above $T_{\rm N1}$. The entropy corresponding to full doublet ground state is recovered at temperature~$\sim$32~K (31~K).\\ 
	The Kondo behavior of resistivity together with the reduced value of entropy at the magnetic transition temperature $T_{\rm m}$ imply a partial quenching of the $4f$\--derived Ce magnetic moment by the Kondo interaction. In such cases the degeneracy of the ground state doublet is partially removed by the Kondo effect, and it has been shown that $S_{\rm 4f}$($T_{\rm m}$)~=~$S_K$($T_{\rm m}$/$T_{\rm K}$), where $S_{\rm 4f}$ is the entropy associated with the magnetic ordering and $S_K$ is the entropy at $T_{\rm m}$ due to the Kondo effect with the Kondo temperature of $T_{\rm K}$~\cite{Mori}. The specific heat and the entropy as a function of T/$T_{\rm K}$ for a spin 1/2 Kondo impurity is known~\cite{Desgranges}, and the ratio $T_{\rm m}$/$T_{\rm K}$ can be determined using the value of $S_{\rm 4f}$. Using this procedure we get a single-ion Kondo temperature $T_{\rm K}$ of 14 and 10~K in CeRhAl$_4$Si$_2$ and CeIrAl$_4$Si$_2$, respectively. On the other hand if $T_{\rm m}$ is taken as the temperature at which the upturn in the heat capacity begins than $T_{\rm K}$ of 12.3 and 7.7~K are obtained. We would like to emphasize that this procedures assumes negligible short range order above the magnetic transition and may therefore overestimate the Kondo temperature. The Kondo temperature of a lattice is believed to be lower than its value for a single impurity. The simple analysis presented here supports the picture of two magnetically ordered compounds with a residual weak Kondo interaction with a Kondo temperature lower or comparable (in the case of Rh-analog) to the magnetic ordering temperature. It may also be noticed that the $T_{\rm K}$ of the Ir compound is lower than that of the Rh analog. This suggests larger electron correlation effect in CeRhAl$_4$Si$_2$, which is amply confirmed by the low temperature heat capacity data discussed below.
	
	The heat capacity below 1~K down to 100~mK, plotted as $C/T$ versus T$^2$ is shown in the insets of Figs.~\ref{HC1}(a) and (b). An extrapolation of the data to $T$~=~0~K gives $\gamma$~=~195.6 and 49.4~mJ/mol K$^2$ in the Rh and its Ir-sibling. The values of $\gamma$ are an order of magnitude larger than the corresponding values in the La-analog and imply a moderate enhancement of the effective electron masses due to the residual Kondo interaction in these two compounds. The upturn in $C/T$ of CeIrAl$_4$Si$_2$ at very low temperatures is attributed to the nuclear Schottky heat capacity arising from the Ir hyperfine quadrupolar interaction.  
	
The $4f$-derived part of the heat capacity in the paramagnetic state in both CeRh(Ir)Al$_4$Si$_2$ shows a broad hump at around 70-80~K indicating a Schottky anomaly due to the splitting of $2J+1$ degenerate levels. This Schottky anomaly is discussed in the next section.

\subsection{Crystalline Electric Field Analysis}

The experimental data presented above show unequivocally that CeTAl$_4$Si$_2$  (T = Rh, Ir) compounds undergo  antiferromagnetic ordering at low temperatures.  Furthermore, the magnetization measurements  revealed a reduced moment of $\sim$~1~$\mu_{\rm B}$/Ce which is much less compared to the free ion value of $g_{\rm J} J (= 6/7~\times 5/2)$, 2.14~$\mu_{\rm B}$/Ce. This fact together with a large negative paramagnetic Weiss temperature $\theta_{\rm p}$,  the negative logarithmic increase of resistivity,  reduced heat capacity jump and a relatively large Sommerfeld coefficient, clearly indicate significant influence of Kondo and crystal electric field (CEF) effects.  To gain more understanding of the magnetic behaviour of these compounds, we have performed a CEF analysis on the magnetization and heat capacity data based on a point charge model.  The Ce-atom in CeTAl$_4$Si$_2$ occupies the $1b$-Wyckoff's position and hence possesses the $4/mmm$ ($\mathcal{D}_{\rm 4h}$) tetragonal point symmetry. For half integral spin of $J (=5/2)$, the crystal field potential will split the $2J+1 (=6)$ degenerate levels into three doublets.  The CEF Hamiltonian for the Ce-atom in a  tetragonal site symmetry is given by,

\begin{equation}
\label{CEF_Ham}
\mathcal{H} = B_2^0O_2^0 + B_4^0O_4^0 + B_4^4O_4^4,
\end{equation}

where $B_l^m$ and $O_l^m$ are the CEF parameters and the Stevens operators, respectively~\cite{Hutchings, Stevens}.  Here we have ignored the 6-th order Stevens operators in the Hamiltonian as they are zero for $J=5/2$.  

In the molecular field approximation, the total susceptibility $\chi$ is expressed in terms of the CEF susceptibility $\chi_{\rm CEFi}$ and the molecular field constant $\lambda$ as, 

\begin{equation}
\label{Chi_CEF}
\frac{1}{\chi_i} = \frac{1}{\chi_{\rm CEFi}} - \lambda_i.
\end{equation}

\begin{figure*}[!]
\includegraphics[width=0.8\textwidth]{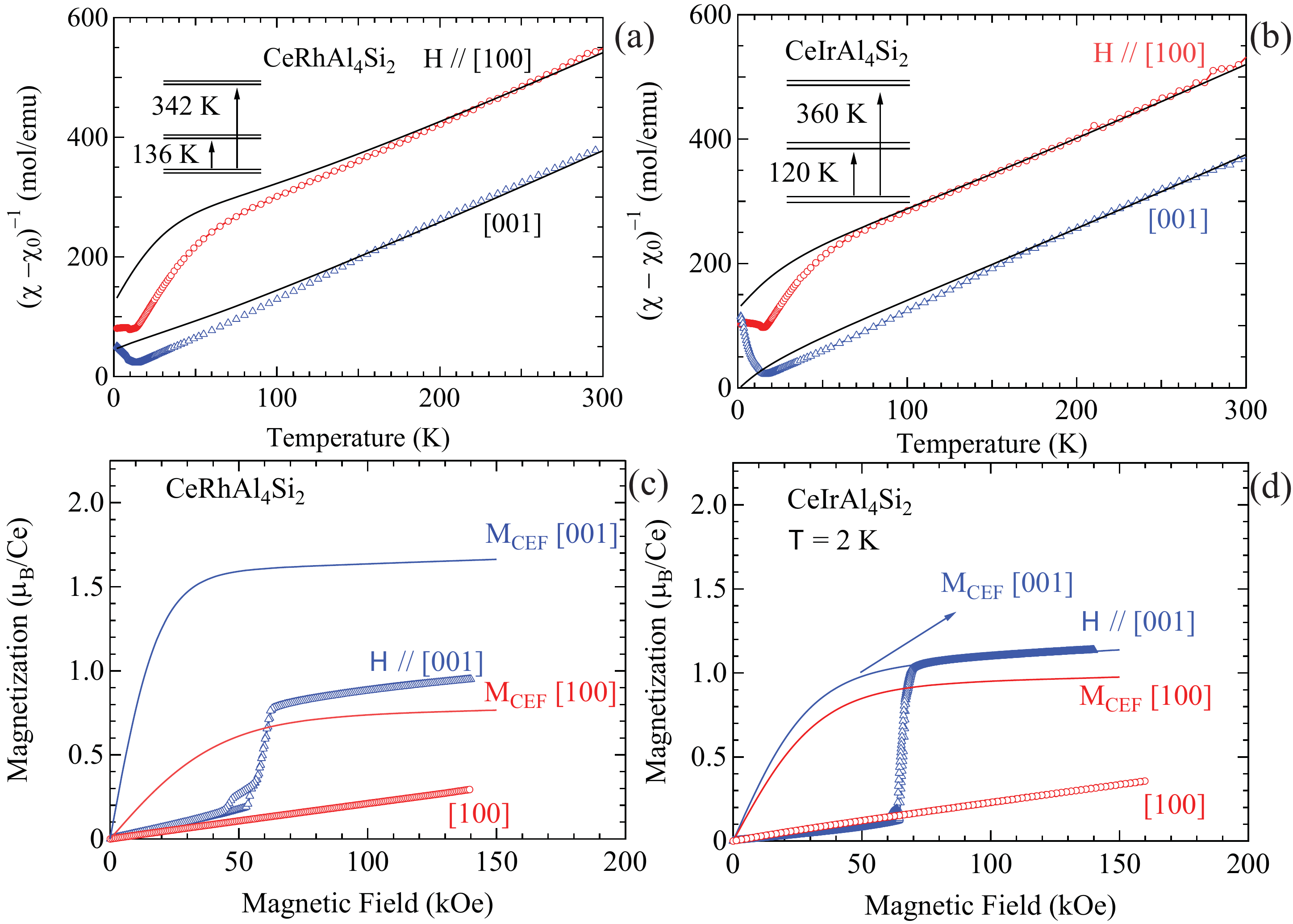}
\caption{\label{fig111}(Color online) (a) and (b) Inverse susceptibility plot of CeRhAl$_4$Si$_2$ and CeIrAl$_4$Si$_2$ and the calculated CEF susceptibility.  The CEF energy levels are also shown.  (c) and (d) Isothermal magnetization measure at $T$~=~2~K and the calculated magnetization curve based on the CEF model.}
\end{figure*}

The  $\chi_{\rm CEFi}$ is given by the standard susceptibility expression, by the combination of  Curie term and the Vanvleck term~\cite{Das}.  It is to be mentioned here that the effective moment $\mu_{\rm eff}$ along both the directions are slightly deviated from the standard value of 2.54~$\mu_{\rm B}$/Ce, for a cerium atom in its trivalent state.  Hence, while performing CEF analysis on the magnetic susceptibility data, we have plotted the inverse susceptibility plot in the form of $1/({\chi - \chi_0})$, where $\chi_0$ is determined from the modified Curie-Weiss law: $\chi = \chi_0 + C/(T-\theta_{\rm p})$ by fixing the effective magnetic moment to 2.54~$\mu_{\rm B}$/Ce.  A similar kind of approach has been made on previous occasions as well~\cite{Thamizh, Takeuchi1}.  Figs.~\ref{fig111}(a) and (b) show the calculated CEF magnetic susceptibility which  explains the anisotropy reasonably well at high temperatures but deviates from the experimental data at low temperatures. Similar quality of fits has earlier been seen, for example, in CeCu$_2$ and CeAgSb$_2$~\cite{Satoh_CeCu2,Takeuchi_CeAgSb2}.   Although it is possible to achieve a better fit to the experimental data, we have chosen the crystal field parameters in such a way that the corresponding crystal field energy levels obtained by diagonalizing the crystal field Hamiltonian (Eq.~\ref{CEF_Ham}) explain the Schottky heat capacity to be discussed later. The crystal field parameters, molecular field constants and the crystal field split energy levels are listed in Table~\ref{Tab_CEF}.  It is to be noted here that the molecular field constant $\lambda$ is very large and highly anisotropic similar to that of the paramagnetic Weiss temperature $\theta_{\rm p}$ as mentioned earlier.  This large value of the molecular field $\lambda$ may be attributed to anisotropic magnetic exchange interactions and due to Kondo effect. Large values of $\lambda$ have, for example, been reported in Kondo antiferromagnets CeCu$_2$ ($\lambda$ = 70 mol/emu), CeIrIn$_5$ ($\lambda$ = -62 mol/emu), CeRhIn$_5$ ($\lambda$ = -36 mol/emu) and CeAgSb$_2$ ($\lambda_{[100]}$ = -28 mol/emu)~\cite{Satoh_CeCu2,Takeuchi_CeAgSb2,Takeuchi_CeTIn5}. The negative value of $\lambda_i$ in CeTAl$_4$Si$_2$ implies the antifrromagnetic interaction of the Ce-4f moments in these compounds. 
%
%****************************Table3******************************
\begin{table}
\caption{\label{Tab_CEF}Crystal field parameters, molecular field constant and the energy levels for CeTAl$_4$Si$_2$ (T = Rh and Ir) obtained from the CEF fitting to the inverse susceptibility plot}
\begin{ruledtabular}
\begin{tabular}{lcc}
%\hline \hline \\
                                & CeRhAl$_4$Si$_2$ & CeIrAl$_4$Si$_2$ \\
                                \hline \\
$B_2^0$ (K)                     & -6.52            & -1.40            \\
$B_4^0$ (K)                     & -0.63            & -0.43            \\
$B_4^4$ (K)                     & 4.74             & 6.34             \\
$\lambda_{\rm [100]}$ (mol/emu) & -122             & -126             \\
$\lambda_{\rm [001]}$ (mol/emu) & -45              & -5               \\
$\Delta_1$ (K)                  & 136              & 120              \\
$\Delta_2$ (K)                  & 342              & 361             \\
%\hline
\end{tabular}
\end{ruledtabular}
\end{table}
%****************************************************************
%
%*********************FIGURE 9***********************************
%****************************************************************

We have also calculated the isothermal magnetization based on the CEF model with the following Hamiltonian:

\begin{equation}
\label{CEF_Magnzn}
\mathcal{H} = \mathcal{H}_{\rm CEF} - g_{\rm J} \mu_{\rm B} J_{i} H,
\end{equation}

where $\mathcal{H}_{\rm CEF}$ is given by Eq.~\ref{CEF_Ham},  the second term is the Zeeman term.  The magnetization $M_i$ is given by the following expression:

\begin{equation}
\label{Magnzn}
M_i = g_J \mu_{\rm B}\sum_n \vert\langle n\vert J_i \vert n\rangle\vert \frac{{\rm exp}(-\beta E_n)}{Z},~~~~(i = x, y, z).
\end{equation}

The calculated CEF magnetization is shown in Fig.~\ref{fig111}(c) and (d) as solid lines.  It is obvious from the figure that quantitatively the calculated CEF magnetization matches poorly with the experimental plots, which is tentatively attributed to the presence of Kondo interaction in these compounds. However, the calculated magnetization reflects the experimentally observed anisotropy.  In the case of the rare-earth atom occupying the tetragonal site symmetry, the sign of the $B_2^0$ parameter usually determines the easy axis or easy plane of magnetization~\cite{Loidl}.  Here the sign of $B_2^0$ is negative which indicates that [001] direction is easy axis of magnetization which is consistent with our experimental data.  

Another estimate of the Kondo temperature has been obtained from the magnetic part of the heat capacity. In a typical Kondo lattice system, the magnetic part of the heat capacity can be thought to be the combination of the Kondo and Schottky contributions, which can be expressed as :

\begin{equation}
\label{Total_4f_HC}
C_{\rm 4f} =  C_K + C_{Sch}.
\end{equation}  

The expression for $C_K$ is given by Schotte and Schotte~\cite{Schotte2} where they assumed a Lorentzian like density of states at the Fermi energy for the impurity spins $S = 1/2$:

\begin{equation}
\label{Kondo_HC}
C_K = k_{\rm B}\frac{\Delta}{\pi k_{\rm B}T} \left(1 - \frac{\Delta}{2 \pi k_{\rm B}T} \psi' \left(\frac{1}{2} + \frac{\Delta}{2 \pi k_{\rm B}T} \right) \right),
\end{equation}

where $\psi'$ is the first derivative of digamma function. Here $\Delta$ is the width of the Lorentzian and is assumed to be  approximately the size of the Kondo energy $k_{\rm B}T_K$. For a three level system $C_{Sch}$ is given by the expression:

\begin{widetext}
\begin{equation}
\label{eqn6}
C_{\rm Sch}= \left[\frac{R}{(k_{\rm B}T)^2} \frac{e^{(\Delta_1 + \Delta_2)/k_{\rm B}T}[-2 \Delta_1 \Delta_2 + \Delta_2^2 (1 + e^{\Delta_1/k_{\rm B}T}) + \Delta_1^2 (1 + e^{\Delta_2/k_{\rm B}T})]}{(e^{\Delta_1/k_{\rm B}T} + e^{\Delta_2 / k_{\rm B}T} + e^{(\Delta_1 + \Delta_2)/k_{\rm B}T})^2} \right],
\end{equation}
\end{widetext}

where $R$ is the universal gas constant and $\Delta_1$ and $\Delta_2$ are the crystal field split excited energy levels. Using the energy levels obtained from the point charge model  of the susceptibility data we have analyzed the magnetic part of the heat capacity using Eq.~\ref{Total_4f_HC}.
%*********************FIGURE 10**********************************
\begin{figure}[!]
\includegraphics[width=0.45\textwidth]{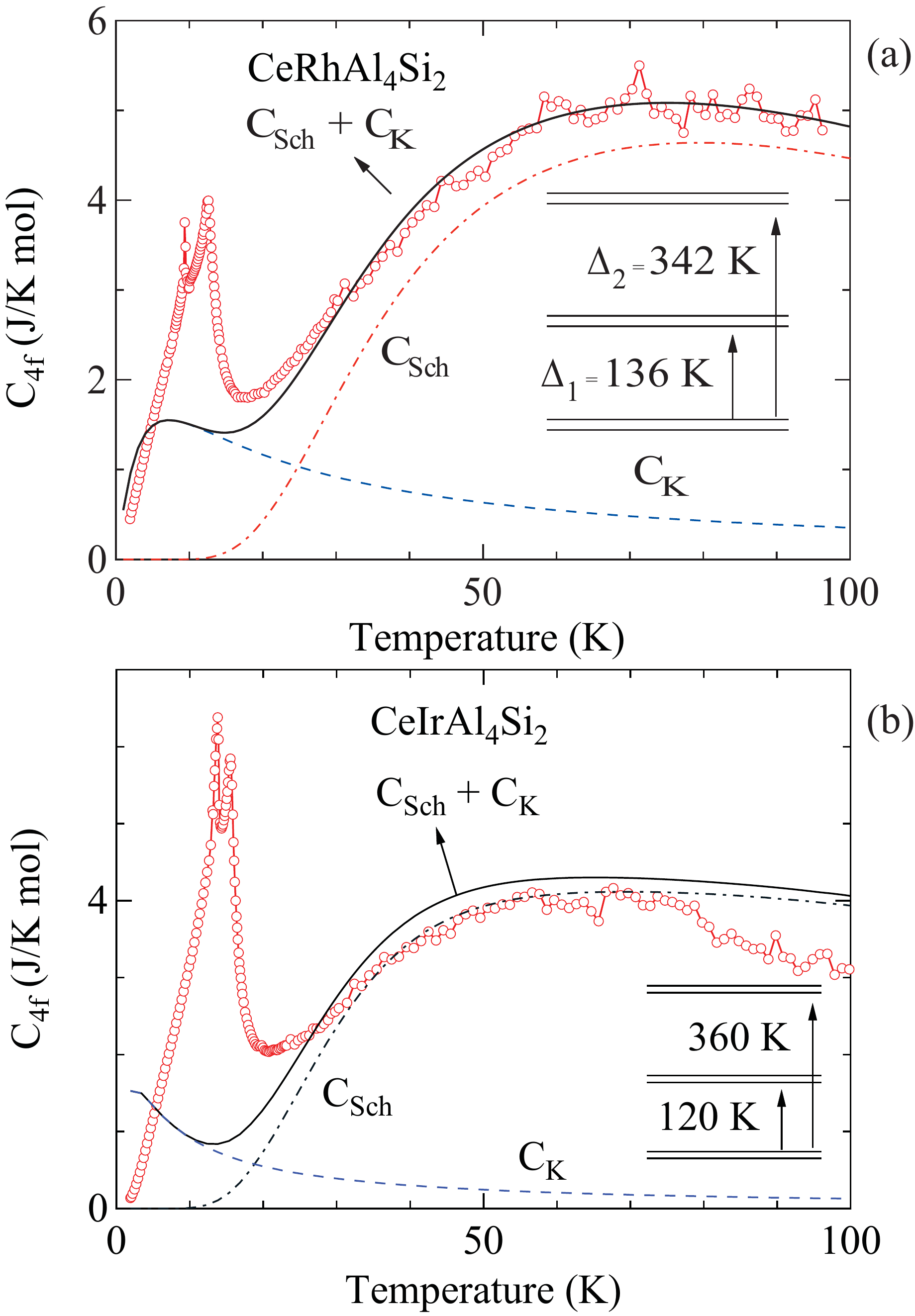}
\caption{\label{fig222}(Color online) The $4f$ contribution to the heat capacity $C_{\rm 4f}$. The thick solid line is sum of the Kondo and Schottky contribution, dashed line is the Kondo contribution $C_{\rm K}$ and dashed-dotted line is due to the Schottky contribution $C_{\rm Sch}$.  The Schottky energy levels are also shown.  }
\end{figure}
%****************************************************************

The solid line in Fig.~\ref{fig222} shows the combined contribution from the Kondo and the Schottky heat capacity, where the two contributions are calculated using Eq.~\ref{eqn6} and Eq.~\ref{Kondo_HC}, respectively. It is evident from the figure that the $4f$-derived part of the heat capacity is reasonably well explained by Eq.~\ref{Total_4f_HC}.  The Kondo temperature $T_{\rm K}$ thus obtained is 10-15~K for CeRhAl$_4$Si$_2$ and 5~K for CeIrAl$_4$Si$_2$, which are comparable to the values obtained using the procedure of Mori \textit{et al.}~\cite{Mori}.  A larger value of $T_{\rm K}$ implies a reduced jump in the heat capacity.  If we refer to Fig.~\ref{fig222}, the jump in the $C_{\rm 4f}$ at the magnetic ordering temperature is smaller for CeRhAl$_4$Si$_2$ compared to that of CeIrAl$_4$Si$_2$ which has a lower $T_{\rm K}$.   Our crystal field calculations on the magnetic susceptibility and heat capacity data indicate that CeTAl$_4$Si$_2$ (T = Rh and Ir) are Kondo lattice systems. 

\subsection{Electronic Structure} 
The local spin density approximation including Hubbard $U$\--electron-correlation–~(LSDA+$U$)~\cite{Anisimov}  approach has been employed to investigate the electronic structure and magnetic properties of CeTAl$_4$Si$_2$ (T~=~Rh, Ir and Pt) compounds.  Calculations have been performed using the scalar relativistic version (which includes the mass velocity and Darwin correction terms) of the LSDA+$U$ method implemented in the tight binding linear muffin tin orbital (TB-LMTO)~\cite{Anderson} and full potential linear augmented plane wave (FP-LAPW)~\cite{Blaha}  methods.  We used $U=6.7$~eV and $J=0.7$~eV to model the onsite $4f$-electron correlations of Ce atoms.  Because of $U$ the occupied spin up states move towards lower energy and the unoccupied spin down states move toward higher energy correcting the positioning of spin up occupied and spin down unoccupied $4f$ states that appear at the Fermi level from the bare LSDA calculations alone. The k-space integrations have been performed with $32\times 32\times 32$ Brillouin zone mesh which was sufficient for the convergence of the total energies (with accuracy of $\sim$~0.1 meV/cell) and magnetic moments (with accuracy of 0.01~$\mu_{\rm B}$/cell).  The experimentally determined $P4/mmm$  crystal structure and lattice constants were used in these calculations. 

%*********************FIGURE 11**********************************
\begin{figure*}[!]
\includegraphics[width=0.95\textwidth]{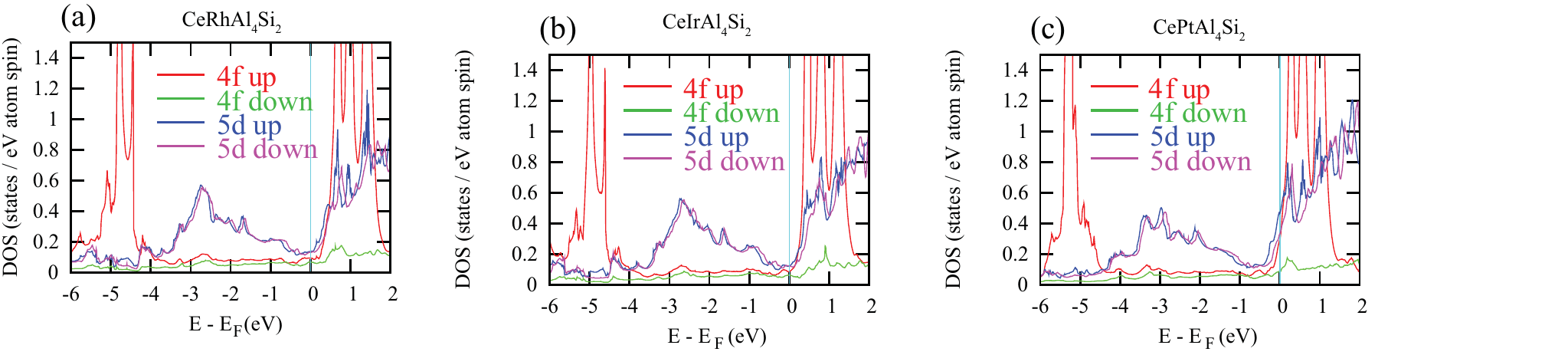}
\caption{\label{Band_Structure}(Color online) $4f$ and $5d$ density of states (DOS) of Ce atoms in CeTAl$_4$Si$_2$ (T~=~Rh, Ir, and Pt).}.
\end{figure*}
%****************************************************************	

Fig.~\ref{Band_Structure} shows $4f$ and $5d$ density of states (DOS) of Ce atoms in CeTAl$_4$Si$_2$ (T~=~Rh, Ir, and Pt).  The occupied spin up $4f$ DOS indicates two peaks and the unoccupied spin up DOS shows three peaks in both CeRhAl$_4$Si$_2$ (Fig.~\ref{Band_Structure}a) and CeIrAl$_4$Si$_2$ (Fig.~\ref{Band_Structure}b).  However, the occupied spin up $4f$ states form single DOS peak in CePtAl$_4$Si$_2$ (Fig.~\ref{Band_Structure}c).  Both $4f$ and $5d$ DOS curves rigidly shift towards the lower energy in CeIrAl$_4$Si$_2$ compared to CeRhAl$_4$Si$_2$.  The rigid shift further increases in CePtAl$_4$Si$_2$   There is no spin polarization between the spin up and spin down $5d$ states in CeRhAl$_4$Si$_2$ (Fig.~\ref{Band_Structure}a) and CeIrAl$_4$Si$_2$ (Fig.~\ref{Band_Structure}b) which is possible only when $4f$ moments at two neighbouring Ce sites are antiparallel. However the $5d$ DOS in CePtAl$_4$Si$_2$ shows spin polarization between spin up and spin down $5d$ states (with ~0.1~$\mu_{\rm B}$ of Ce 5d) which is possible when $4f$ moments at two neighbouring Ce sites are parallel~\cite{Ghimire}. All the three compounds CeRhAl$_4$Si$_2$, CeIrAl$_4$Si$_2$, and CePtAl$_4$Si$_2$ compounds show identical $4f$ moment (1.1~$\mu_{\rm B}$), which is comparable to  the experimentally observed $4f$ local moment in these compounds.   Furthermore, our calculations also indicate similar complex quasi\--2D and nesting features of the Fermi surfaces in CeTAl$_4$Si$_2$ compounds as obtained in LaTAl$_4$Si$_2$ counterparts (Ref.~\onlinecite{Ghimire}). Since the Ce $4f$-states are away from the Fermi level, it is not surprising to obtain similar Fermi surface features in CeTAl$_4$Si$_2$ and the corresponding LaTAl$_4$Si$_2$ compounds. However, when treating Ce-4f as delocalised bands, we find Fermi surface features very similar to LaPtAl$_4$Si$_2$ as pointed earlier in ref.~\onlinecite{Ghimire}. We note here that similar complex Fermi surface features and enhanced electronic coefficient of the specific heat have also been observed in Ce$_2$PdGe$_3$~\cite{Baumbach}. Interestingly the total DOS at the Fermi level is higher by ~19\% in CeRhAl$_4$Si$_2$ compared to CeIrAl$_4$Si$_2$, which indicates a good trend with the observed coefficients of the electronic specific heat. As pointed out in ref.~\onlinecite{Baumbach}, the shapes of the Fermi surfaces may be reasonable but the mass renormalization, which is a dynamical entity, may not be fully captured in the electronic structure calculation.   

The origin of the unusual physical properties, such as Kondo and heavy fermion behaviour, of some cerium compounds.~\cite{Pickett,Jarlborg} is believed to be due to an unusual $4f$ density of states peak at the Fermi level~\cite{Jarlborg_CeAl2} depending on the onsite electron correlations. The calculations with reduced Hubbard $U$ parameter (e.g., for $U~=~3$~eV) indicate that the double peak $4f$ DOS in CeRhAl$_4$Si$_2$ and CeIrAl$_4$Si$_2$ and single peak $4f$ DOS in CePtAl$_4$Si$_2$ (seen with $U=6.7$~eV) shift towards the Fermi level (without significantly affecting the unoccupied $4f$ DOS) altering their coupling to the conduction electrons close to the Fermi level.  Due to the formation of two $4f$ DOS peaks and low Ce$5d$ DOS in CeRhAl$_4$Si$_2$ and CeIrAl$_4$Si$_2$ there is a relatively weak hybridization between the $4f$ and the conduction electron states including  Rh$4d$/Ir$5d$  at the Fermi level. On the other hand, the formation of single $4f$ DOS peak and higher Ce $4f$ and $5d$ DOS in CePtAl$_4$Si$_2$ results in a slightly stronger hybridization between the $4f$ and the conduction electron states including Pt $5d$ at the Fermi level.

\section{Conclusion}

We have successfully synthesized single crystals of CeTAl$_4$Si$_2$ (T = Rh and Ir)  by using the Al-Si binary eutectic composition as flux.  The grown crystals  have platelet like morphology with (001)-plane perpendicular to the plane of the crystal.  Our comprehensive thermal and transport studies reveal that both compounds order antiferromagnetically with two N\'{e}el temperatures.  A sharp metamagnetic transition is observed for $H~\parallel$~[001] direction where as the magnetization is relatively small and varies linearly for $H~\parallel$~[100] direction, indicating that [001] direction is the easy axis of magnetization in both the cases, which is consistent with neutron diffraction. A large negative paramagnetic Weiss temperature $\theta_{\rm p}$, a reduced magnetization, a reduced magnetic entropy and a negative logarithmic increase in the resistivity with decreasing temperature indicate that these compounds are Kondo lattice systems. The antiferromagnetic transition temperature decreases with pressure, confirming the presence of Kondo interaction  in the two compounds. We have performed detailed crystal electric field calculations on the magnetization and heat capacity data and estimated the Kondo temperature and the energy levels of the $2J+1$ degenerate ground state. It is found from our analysis that exchange interaction is anisotropic and the Kondo temperature is larger for Rh system. Our detailed electronic structure calculations provide an insight into the observed  magnetic and Kondo anomalies of these systems.

The electronic structure part of the work was supported by the U.S. Department of Energy, Office of Basic Energy Science, Division of Materials Sciences and Engineering. The Ames Laboratory is operated for the U.S. Department of Energy by Iowa State University under Contract No. DE-AC02-07CH11358.


\begin{thebibliography}{99}


\bibitem{Maurya} A. Maurya, A. Thamizhavel, A. Provino, M. Pani, P. Manfrinetti, D. Paudyal and S. K. Dhar, Inorg. Chem., {\bf 53}, 1443 (2014).

\bibitem{Maurya2} A. Maurya, P. Bonville, A. Thamizhavel and S. K. Dhar, arXiv:1411.0379v1 (2014).

\bibitem{Wu} X. Wu and M. G. Kanatzidis,  J. Solid State Chem., {\bf 178}, 3233 (2005). 

\bibitem{Latturner} S. E. Latturner and M. G. Kanatzidis,  Inorg. Chem., {\bf 47}, 2089 (2008). 

\bibitem{Sieve} B.Sieve, X. Chen, J. Cowen, P. Larson, S. D. Mahanti and and M. G. Kanatzidis,  Chem. Mater., {\bf 11}, 2451 (1999). 

\bibitem{Stewart} G. R. Stewart, Rev. Mod. Phys., {\bf 56}, 755, (1984).

\bibitem{Steglich}N. Gerwe and F. Steglich, Handbook on the physics and chemistry of rare earths, North-Holland, Amsterdam {\bf 14}, p. 343 (1991).

\bibitem{Stewart2} G. R. Stewart, Rev. Mod. Phys., {\bf 73}, 4, (2001).

\bibitem{Ghimire}N. J. Ghimire, F. Ronning, D. J. Williams, B. L. Scott, Y. Luo, J. D. Thompson and E. D. Bauer,J. Phys.: Condens. Matter, {\bf 27}, 025601 (2015).

\bibitem{Doniach} S. Doniach, Physica B+C {\bf 91}, 231 (1977).

\bibitem{Ghimire_Neutron}N. J. Ghimire, S. Calder, M. Janoschek and E. D. Bauer, arXiv:1503.06860 (2015)


\bibitem{Carvajal} Juan Rodriguez-Carvajal, Physica B {\bf 192}, 55 (1993).

\bibitem{Thompson} J. D. Thompson 1993 \textit{Selected Topics in Magnetism} \textbf{Vol.2} edited by L. C. Gupta and M. S. Multani (World Scientific) p.107.

\bibitem{Maurya_EuNiGe3}A. Maurya, P. Bonville, A. Thamizhavel and S.K. Dhar, J. Phys.: Condens. Matter {\bf 26}, 216001 (2014).

\bibitem{Fournier} J.M. Fournier and E. Gratz 1993 \textit{Handbook on the Physics and Chemistry of rare earths: Transport properties of rare earth and actinide intermetallics} \textbf{vol.17} eds. K.A. Gschneidner, Jr. L. Eyring, G.H. Lander and G.R. Choppin (Elsevier Science Publishers) p409.

\bibitem{Zlatic}V. Zlatic, J. Phys. F {\bf11}, 2147 (1981).

\bibitem{Jan}J.-P. Jan in Solid Stata Physics, vol.5, eds. F. Seitz and D. Turnbull (Academic Press Inc. New York 1957) p1.

\bibitem{Mori} H. Mori, H. Yashima and N. Sato, J. low. temp. phys. {\bf 58}, 513, (1985).

\bibitem{Desgranges} H. -U. Desgranges and K. D. Schotte, Phys. Lett. {\bf 91A}, 240, {1942}.

\bibitem{Hutchings} M. T. Hutchings 1965 \textit{Solid State Physics: Advances in Research and Applications} \textbf{Vol.16} edited by F. Seitz and B. Turnbull (New York: Academic) p.227.

\bibitem{Stevens} K. W. H. Stevens  \textit{Proc. Phys. Soc.} (London)  \textbf{Sect.A65}, 209 (1952).

\bibitem{Das}P. K. Das, N. Kumar, R. Kulkarni and A. Thamizhavel, Phys. Rev. B, {\bf 83}, 134416 (2011).

\bibitem{Thamizh}A. Thamizhavel, R. Kulkarni and S. K. Dhar, Phys. Rev. B, {\bf 75}, 144426 (2007).

\bibitem{Takeuchi1}T. Takeuchi, S. Hashimoto, T. Yasuda, H. Shishido, T. Ueda, M. Yamada, Y. Obiraki, M. Shiimoto, H. Kohara, T. Yamamoto, K. Sugiyama, K. Kindo, T. D. Matsuda, Y. Haga, Y. Aoki, H. Sato, R. Settai and Y.\={O}nuki, J. Phys.: Condens. Matter, {\bf 16}, L33 (2004).

\bibitem{Satoh_CeCu2} K. Satoh, A. Fukada, I. Umehara, Y. \={O}nuki, H. Sato, and S. Takayanagi, J. Phys. Soc. Jpn. {\bf61}, 3267 (1992).

\bibitem{Takeuchi_CeAgSb2}T. Takeuchi, A. Thamizhavel, T. Okubo, M. Yamada, N. Nakamura, T. Yamamoto, Y. Inada, K. Sugiyama, A. Galatanu, E. Yamamoto, K. Kindo, T. Ebihara, and Y. \={O}nuki, Phys. Rev. B {\bf67}, 064403 (2003).

\bibitem{Takeuchi_CeTIn5} T. Takeuchi, T. Inoue, K. Sugiyama, D. Aoki, Y. Tokiwa, Y. Haga, K. Kindo, and Y. \={O}nuki, J. Phys. Soc. Jpn. {\bf70}, 877 (2001).

\bibitem{Loidl}A. Loidl, K. Knorr, G. Knopp, A. Krimmel, R. Caspary, A. B\"{o}hm, G. Sparn, C. Geibel, F. Steglich and A. P. Murani, Phys. Rev. B {\bf 46}, 9341 (1992).

\bibitem{Schotte2}K. D. Schotte and U. Schotte, Phys. Lett. {\bf 55A}, 38, (1975).


%\bibitem{Jensen}J. Jensen and A. R. Mackintosh, Rare earth magnetism structures and excitations, Calrendon press, Oxford {\bf Chap 2}, p. 73 (1991).

\bibitem{Anisimov}V. I. Anisimov, F. Aryasetiawan, and A.I. Lichtenstein, J. Phys.: Condens. Matter. {\bf 9}, 767 (1997).

\bibitem{Anderson}O. K. Andersen and O. Jepsen, Phys. Rev. Lett. {\bf 53}, 2571 (1984).

\bibitem{Blaha}P. Blaha, K. Schwarz, G. Madsen, D. Kvasnicka, and J. Luitz, \textit{WIEN2k}, An Augmented Plane Wave + Local Orbitals Program for Calculating Crystal Properties (Karlheinz Schwarz, Techn. Universität Wien, Austria), ISBN 3-9501031-1-2 (2001).  


\bibitem{Baumbach}R.E. Baumbach, A. Gallagher, T. Besara, J. Sun, T. Siegrist, D.J. Singh, J.D. Thompson, F. Ronning and E.D. Bauer, Phys. Rev. B {\bf91}, 035102 (2015).

\bibitem{Pickett}W. E. Pickett  and B. M. Klein, Journal of the Less-Common Metals, {\bf 93}, 219 (1983).

\bibitem{Jarlborg}T. Jarlborg, A. J. Freeman, and D. D. Koelling, J. Magn. Magn. Mater {\bf 60}, 291 (1986).


\bibitem{Jarlborg_CeAl2}T. Jarlborg, A. J. Freeman, and D. D. Koelling, J. App. Phys. {\bf 53}, 2140 (1982).

%\bibitem{Fulde}P. Fulde, J. Keller, and G. Zwicknagl in Theory of Heavy fermion Systems, Solid State Physics Vol. {\bf 41}, page (1-150).



\end{thebibliography}
\end{document}